\documentclass{article}
\usepackage[a4paper, total={5.5in, 8in}]{geometry}
\usepackage{graphicx} 
\usepackage{tikz}
\usepackage{amsmath}
\usepackage{amssymb}
\usepackage{multirow}
\usepackage{authblk}
\usepackage{hyperref}

\newtheorem{theorem}{Theorem}
\newtheorem{proposition}{Proposition}

\newtheorem{lemma}{Lemma}

\newcommand{\R}{\mathbb{R}}

\newcommand{\PP}{{\mathbb{P}}}
\newcommand{\E}{\mathbf{E}}
\newcommand{\xnew}{x_{*}}
\newcommand{\argmin}{\arg\min}
\newcommand{\G}{\mathrm{G}}

\usepackage{natbib}
\title{Statistical Inference in High-dimensional Poisson Regression with Applications to Mediation Analysis}

\author[1]{Prabrisha Rakshit}
\author[2]{Zijian Guo}
\affil[1]{University of Pennsylvania}
\affil[2]{Rutgers, The State University of New Jersey}

\begin{document}

\maketitle

\begin{abstract}
\noindent Large-scale datasets with count outcome variables are widely present in various applications, and the Poisson regression model is among the most popular models for handling count outcomes. This paper considers the high-dimensional sparse Poisson regression model and proposes bias-corrected estimators for both linear and quadratic transformations of high-dimensional regression vectors. We establish the asymptotic normality of the estimators, construct asymptotically valid confidence intervals, and conduct related hypothesis testing. We apply the devised methodology to high-dimensional mediation analysis with count outcome, with particular application of  testing for the existence of interaction between the treatment variable and high-dimensional mediators. We demonstrate the proposed methods through extensive simulation studies and application to real-world epigenetic data.
\end{abstract}

\section{Introduction}
\label{sec: intro}
Modern data analysis often deals with datasets with many potential predictors compared to the number of observations, making high-dimensional regression an essential statistical tool. High-dimensional regression finds applications in various fields, such as genetics, economics, and social sciences. Moreover, in addition to continuous or binary outcomes, outcome variables are often count variables, such as the number of car accidents, calls to a customer service center, or defects in a manufacturing process.

\vspace{2 mm}

\noindent This paper focuses on the statistical inference in the high-dimensional Poisson regression model and discusses its application to the high-dimensional mediation analysis in Section \ref{sec: mediation}. We consider the high-dimensional Poisson regression model :
\begin{equation}
    y_i \sim \text{Poisson}(e^{X_{i\cdot}^{\intercal}\beta}) \quad \text{for} \quad 1 \leq i \leq n, 
    \label{eq: GLM}
\end{equation}
where $y_i \in \mathbb{Z}$ is a count outcome variable, $X_{i\cdot} \in \mathbb{R}^p$ denotes the high-dimensional covariates and $\beta \in \mathbb{R}^p$ is a high-dimensional sparse regression vector with sparsity $k$. Here we assume that $\{y_i,X_{i\cdot}\}_{1\leq i \leq n}$ are independently and identically distributed.
 
\vspace{2 mm}
\noindent Our first goal is to construct confidence interval (CI) for the linear functional $x_*^{\intercal}\beta$  for arbitrary $x_* \in \mathbb{R}^p$ and conduct the following hypothesis testing
    \begin{equation}
        H_0 : x_*^{\intercal}\beta = 0.
        \label{eq: hypothesis linear}
    \end{equation}
An important motivation for making inference about $ x_*^{\intercal}\beta $ lies in its relevance to the prediction problem in high-dimensional Poisson regression. This prediction problem closely involves the quantity $x_*^{\intercal}\beta$, where $x_* \in \mathbb{R}^p$ represents the covariate values for the individual whose count outcome we aim to predict. This paper systematically addresses the problem of linear functional inference in high-dimensional Poisson regression for the first time.

\vspace{2 mm}
\noindent Next, we consider inference for the quadratic functional $\beta_G^{\intercal}\mathrm{A}\beta_G$ where $G \subset \{1,2,\ldots,p\}$ a given index set and $A\in \R^{|G|\times |G|}$ is positive definite, with $|G|$ denoting the cardinality of $G$. The quadratic form inference is closely related to the group significance test,
\begin{equation}
    H_0 : \beta_G = 0,
    \label{eq: significance}
    \end{equation}
For any positive definite matrix $A \in \mathbb{R}^{|G|\times|G|}$, testing \eqref{eq: significance} is equivalent to testing
    \begin{equation}
        H_{0A} : \beta_G^{\intercal}\mathrm{A}\beta_G = 0.
        \label{eq: hypothesis A}
    \end{equation}
A group significance test can be useful to identify sets of predictor variables that are jointly influential for predicting a count outcome. In this paper, we tackle the group inference problem in high dimensions, particularly for large-sized groups $G$. For the weighting matrix $A$, we particularly focus on the case of known $A$ (e.g.,  $A = {\rm I} $) and $A = \Sigma_{G,G}$, where $\Sigma = \mathbb{E}X_{i\cdot}X_{i\cdot}^{\intercal}$ denotes the unknown population covariance matrix. This paper provides a systematic approach to addressing both linear and quadratic functional inferences within high-dimensional Poisson regression, which, to our knowledge, has not been explored in this manner before.

\vspace{2 mm}
\noindent In section \ref{sec: mediation}, we introduce how to apply the inference method for the high-dimensional Poisson regression to the mediation analysis setup. Introduced by \cite{baronmed}, mediation analysis has been broadly applied in many scientific disciplines, such as sociology, psychology, behavioral science, economics, epidemiology, public health science and genetics \citep{medapp1, medapp2, medapp3, medapp4, medapp5}. Count outcome variables are frequently encountered in the context of mediation analysis, where Poisson regression can be used to explore the indirect effects of the exposure intervened through the mediator. Examples include examining the effect of alcohol advertising on youth drinking behavior (number of drinks consumed), mediated through peer drinking \citep{countmed1}; assessing the relationship between parental education and child health (number of days with a cold), intervened by family income and access to health care \citep{countmed2}; investigating the effect of maternal age on the number of perinatal illnesses mediated by infant birthweight \citep{countmed3}.

\vspace{2 mm}
\noindent Here, we consider a simplified mediation scenario, primarily aiming to demonstrate the application of our inference method rather than making a significant contribution to mediation analysis itself. The method we propose could be combined with other approaches to address more complex and realistic situations that arise in practical settings.

\subsection{Main Results and Contributions}
\label{sec: contribution}

The penalized estimation methods have been well developed to estimate $\beta \in \mathbb{R}^p$ in the high-dimensional Poisson regression 
\citep{buhlmann2011statistics,negahban2009unified,adjlasso,adplasso,elasticnet,measerror}. The penalized estimators enjoy desirable estimation accuracy properties. However, these estimators are not suited for confidence interval construction because the bias introduced by the penalty dominates the total uncertainty. Bias-correction procedure was introduced in \cite{van2014asymptotically,javanmard2014confidence,zhang2014confidence} to perform inference for individual regression coefficient in high-dimensional linear regression. Extensions of this work include inference on linear combinations of coefficients \citep{cai2017confidence, athey2018approximate, zhu2018linear, cai2019individualized}, and recent progress has been made on logistic regression models as well \citep{linlog,GLM}. Yet, Poisson regression poses additional challenges due to its unique structure, such as the count-based nature of the response variable and the highly non-linear link function, making existing methods inadequate.

\vspace{2 mm}

\noindent In this paper, we introduce a framework for conducting inference on both linear and quadratic functionals in high-dimensional Poisson regression, a problem that has not been addressed before. While methods for inference in linear and logistic regression models have been developed, Poisson regression has remained largely unexplored in this context. Our approach builds on advanced bias-correction and debiasing techniques to address these challenges, offering a new direction for inference in high-dimensional Poisson models. Our contributions are twofold:
\begin{enumerate}
    \item We propose a general framework for handling prediction and inference problems in high-dimensional Poisson regression. This includes developing methods for arbitrary linear combinations of regression coefficients and testing the joint significance of predictors through weighted quadratic functionals. Our method uses a weighting scheme that bypasses the difficult inversion of the Hessian matrix of the negative log-likelihood.
    
    \item We extend our framework to high-dimensional mediation analysis involving count outcomes, a domain that has previously focused on continuous and binary outcomes. We apply this approach to a large-scale genomic study on post-traumatic stress disorder (PTSD) \citep{PTSD}, exploring the mediation effect of DNA methylation changes in the relationship between childhood maltreatment and adult PTSD.
\end{enumerate}
To the best of our knowledge this work is the first to address inference for linear and quadratic functionals in high-dimensional Poisson regression, overcoming significant challenges inherent in the model’s complexity.

\subsection{Related Work}
\label{sec: related work}
Our method provides a general framework for the inference in high-dimensional Poisson regression. As mentioned in section \ref{sec: contribution}, inference methods are well studied in the case of linear and logistic regression but much less explored with respect to Poisson regression. However, Poisson regression finds important applications in various domains including mediation analysis, as pointed out in section \ref{sec: intro}.

\vspace{2 mm}

\noindent Additionally, the theoretical guarantee for our method requires much weaker assumptions compared to the existing literature. \cite{sshdiglm} performed post-selection inference in high-dimensional GLM by splitting the sample into two equal halves. The idea is to use the first half for variable selection and the second half with the reduced set of selected variables for statistical inference. \cite{poisonline} developed online estimation and inference methods for single or low-dimensional regression coefficients in high-dimensional GLMs. The theoretical guarantees require the assumption of sure screening property, i.e., it assumes that the selected sub-model contains the true model with high probability. Unfortunately, performing exact variable screening appears impractical in real-world applications. The validation of our method does not rely on such assumptions.

\vspace{2 mm}

\noindent \cite{streaming} introduced an online debiased procedure for high-dimensional GLMs but assumed the invertibility of the Fisher information matrix. We use the re-weighting technique to bypass imposing invertibility or sparsity assumptions on the precision or the Hessian matrix. See section \ref{sec: LF} for details. 

\vspace{2 mm}

\noindent \cite{linearhyp} proposed a partial penalized likelihood ratio test to deal with linear hypothesis testing for high-dimensional GLMs. However it assumes that the minimum absolute value of the nonzero regression coefficients is much larger than a specific threshold. Our theory does not impose any such restriction on the minimum signal strength.

\vspace{2 mm}
\noindent \cite{guo2019optimal} and \cite{geneticrelated} studied inference for $\beta^{\intercal}A\beta$ in case of high-dimensional linear and logistic regressions respectively, which is a special case of \eqref{eq: hypothesis A} by setting $G=\{1,2, \cdots, p\}$. Inference for $\beta_G^{\intercal}A\beta_G$ in high-dimensional linear regression is well-studied in \cite{grouplin}. However, the analogous problem has not yet been explored in Poisson regression, due to the non-linearity of the model and the complex structure of the Hessian matrix.

\subsection{Organization and Notation}
The rest of the paper is organized as follows. 
In section \ref{sec: lin quad inf}, we construct CIs and statistical tests for linear and quadratic functionals of the regression coefficients. We then study in section \ref{sec: theory} the theoretical properties of the proposed estimators and establish their asymptotic normality under suitable sparsity condition. In section \ref{sec: mediation}, we apply the procedure to the interesting problem of high-dimensional mediation analysis with count outcome. The extensive numerical studies are conducted in section \ref{sec: sim} to support our theory. Finally, in section \ref{sec: real data Poisson}, we illustrate the method through the analysis of epigenetic data. Proofs and additional
numerical results are presented in section 7.

\vspace{2 mm}
\noindent We finish this section with notation. Throughout, for a vector $a=\left(a_1, \ldots, a_n\right)^{\top} \in \mathbb{R}^n$, we define the $\ell_p$ norm $\|a\|_p=\left(\sum_{i=1}^n\left|a_i\right|^p\right)^{1 / p}$, the $\ell_0$ norm $\|a\|_0=\sum_{i=1}^n 1\left\{a_i \neq 0\right\}$, and the $\ell_{\infty}$ norm $\|a\|_{\infty}=\max _{1 \leq j \leq n}\left|a_i\right|$, and let $a_{-j} \in \mathbb{R}^{n-1}$ stand for the subvector of $a$ without the $j$-th component. For a matrix $A \in \mathbb{R}^{p \times q}, \lambda_i(A)$ stands for the $i$-th largest singular value of $A$ and $\lambda_{\max }(A)=\lambda_1(A), \lambda_{\min }(A)=\lambda_{\min \{p, q\}}(A)$. 
For a matrix $X\in \R^{n\times p}$, $X_{i\cdot}$,  $X_{\cdot j}$ and $X_{i,j}$  denote respectively the $i$-th row,  $j$-th column, $(i,j)$ entry of the matrix $X$. $X_{i,-j}$ denotes the sub-row of $X_{i\cdot}$ excluding the $j$-th entry. Let $[p]=\{1,2,\cdots,p\}$. For a subset $G\subset[p]$ and a vector $x\in \R^{p}$, $x_{G}$ is the subvector of $x$ with indices in $G$ and $x_{-G}$ is the subvector with indices in $G^{c}$. Similarly for a matrix $X \in \R^{p \times p}$, $X_{G,G}$ is the submatrix of $X$ with rows and columns restricted to $G$. Similarly $X_{iG}$ and $X_{Gj}$ denote the $i-$th row of $X$ with columns in $G$ and $j-$th column of $X$ with rows in $G$ respectively.
For positive sequences $\left\{a_n\right\}$ and $\left\{b_n\right\}$, we write $a_n=o\left(b_n\right), a_n \ll b_n$ or $b_n \gg a_n$ if $\lim _n a_n / b_n=0$, and write $a_n=O\left(b_n\right), a_n \lesssim b_n$ or $b_n \gtrsim a_n$ if there exists a constant $C$ such that $a_n \leq C b_n$ for all $n$. We write $a_n \asymp b_n$ if $a_n \lesssim b_n$ and $a_n \gtrsim b_n$.

\section{Inference for Linear and Quadratic Functionals}
\label{sec: lin quad inf}
In this section, we describe our method of constructing point and interval estimators for $x_*^{\intercal}\beta$ for arbitrary $x_* \in \R^p$ and $\beta_G^{\intercal}A\beta_G$ where $G \subset \{1,2,\ldots,p\}$ and $A \in \R^{|G|\times|G|}$. { We define the following penalized estimator for Poisson regression \citep{buhlmann2011statistics}
\begin{equation}
    \widehat{\beta} = \argmin_{\beta \in \mathbb{R}^p}\ell(\beta) + \lambda\|\beta\|_1,
    \label{eq: lasso}
\end{equation}
with the negative log-likelihood $\ell(\beta) = \frac{1}{n}\sum_{i=1}^{n}\left(e^{X_{i\cdot}^{\intercal}\beta} - y_iX_{i\cdot}^{\intercal}\beta - \log(y_i!)\right)$ and $\lambda \asymp \sqrt{{\log p}/{n}}$.}

\noindent Although the penalized estimator $\widehat{\beta}$ achieves the optimal rate of convergence, the plug-in estimators $x_*^{\intercal}\widehat{\beta}$ and $\widehat{\beta}_G^{\intercal}\widehat{A}\widehat{\beta}_G$ cannot be directly used for constructing confidence intervals. This limitation arises because the bias in these estimators can be as large as their variance. To address this problem, we propose an estimator that corrects the bias in the plug-in estimators by using appropriately chosen weights. The existing literature \citep{van2014asymptotically, javanmard2014confidence, zhang2014confidence} suggests constructing a bias-corrected estimator for $\beta_j$ using the following approach: 
$$
\widehat{\beta}_j + \widehat{u}^{\intercal} \frac{1}{n} \sum_{i=1}^{n} X_{i\cdot} (y_i - e^{X_{i\cdot}^{\intercal}\widehat{\beta}}).
$$
In this expression, $\widehat{u} \in \mathbb{R}^p$ is the projection direction used to correct the bias of $\widehat{\beta}_j$. Additionally, we define the error term \(\epsilon_i\) as follows for \(1 \leq i \leq n\):
$\epsilon_i = y_i - e^{X_{i\cdot}^{\intercal}\beta}$. Using the Taylor expansion,  we have
\begin{eqnarray}
    \frac{1}{n}\sum_{i=1}^{n}X_{i\cdot}\left(y_i - e^{X_{i\cdot}^{\intercal}\widehat{\beta}}\right) & = & \frac{1}{n}\sum_{i=1}^{n}X_{i\cdot}\left(e^{X_{i\cdot}^{\intercal}\beta} - e^{X_{i\cdot}^{\intercal}\widehat{\beta}}\right) + \frac{1}{n}\sum_{i=1}^{n}X_{i\cdot}\epsilon_i \nonumber \\
    & = & -\frac{1}{n}\sum_{i=1}^{n}e^{X_{i\cdot}^{\intercal}\widehat{\beta}}X_{i\cdot}X_{i\cdot}^{\intercal}\left(\widehat{\beta}-\beta\right) + \frac{1}{n}\sum_{i=1}^{n}X_{i\cdot}\Delta_i + \frac{1}{n}\sum_{i=1}^{n}X_{i\cdot}\epsilon_i,
\end{eqnarray}
where $\Delta_i = e^{X_{i\cdot}^{\intercal}\widehat{\beta} + tX_{i\cdot}^{\intercal}(\beta - \widehat{\beta})}\left(X_{i\cdot}^{\intercal}(\widehat{\beta}-\beta)\right)^{2}$ is the second order remainder term from the Taylor series expansion. Therefore to estimate the error $\widehat{\beta} - \beta$ we will have to handle the complicated matrix $\frac{1}{n}\sum_{i=1}^{n}e^{X_{i\cdot}^{\intercal}\widehat{\beta}}X_{i\cdot}X_{i\cdot}^{\intercal}$, the Hessian matrix of the negative log-likelihood $\ell(\beta)$. Handling this Hessian matrix is difficult owing to its dependence on $\widehat{\beta}$. In order to solve this issue, we perform the following weighted decomposition :
\begin{equation}
    \frac{1}{n}\sum_{i=1}^{n}e^{-X_{i\cdot}^{\intercal}\widehat{\beta}}X_{i\cdot}\left(y_i - e^{X_{i\cdot}^{\intercal}\widehat{\beta}}\right) = \widehat{\Sigma}\left(\beta-\widehat{\beta}\right) + \frac{1}{n}\sum_{i=1}^{n}e^{-X_{i\cdot}^{\intercal}\widehat{\beta}}X_{i\cdot}\Delta_i + \frac{1}{n}\sum_{i=1}^{n}e^{-X_{i\cdot}^{\intercal}\widehat{\beta}}X_{i\cdot}\epsilon_i,
    \label{eq: weight decomp}
\end{equation}
where $\widehat{\Sigma} = \frac{1}{n}X^{\intercal}X$. The advantage of the weighted decomposition is that we now need to deal with the sample covariance matrix $\widehat{\Sigma}$ instead of $\frac{1}{n}\sum_{i=1}^{n}e^{X_{i\cdot}^{\intercal}\widehat{\beta}}X_{i\cdot}X_{i\cdot}^{\intercal}$ to estimate the vector difference $\widehat{\beta} - \beta$.

\subsection{Inference for Linear Functional}
\label{sec: LF}
We propose the following estimator for $x_*^{\intercal}\beta$,
\begin{equation}
    \widehat{x_*^{\intercal}\beta} = x_*^{\intercal}\widehat{\beta} + \widehat{u}^{\intercal}\frac{1}{n}\sum_{i=1}^{n}e^{-X_{i\cdot}^{\intercal}\widehat{\beta}}X_{i\cdot}\left(y_i - e^{X_{i\cdot}^{\intercal}\widehat{\beta}}\right).
    \label{eq: correction}
\end{equation}
To construct the projection direction $\widehat{u}$, we decompose the error $\widehat{x_*^{\intercal}\beta} - x_*^{\intercal}\beta$ using \eqref{eq: weight decomp},
\begin{equation}
    \widehat{x_*^{\intercal}\beta} - x_*^{\intercal}\beta = \widehat{u}^{\intercal}\frac{1}{n}\sum_{i=1}^{n}e^{-X_{i\cdot}^{\intercal}\widehat{\beta}}X_{i\cdot}\epsilon_i + \left(x_* - \widehat{\Sigma}\widehat{u}\right)^{\intercal}(\widehat{\beta}-\beta) + \widehat{u}^{\intercal}\frac{1}{n}\sum_{i=1}^{n}e^{-X_{i\cdot}^{\intercal}\widehat{\beta}}X_{i\cdot}\Delta_i.
    \label{eq: error decomposition}
\end{equation}
\noindent Motivated by the above decomposition we construct $\widehat{u}$ as the solution to the following optimization problem
\begin{eqnarray}
\widehat{u}=\;\argmin_{u\in \mathbb{R}^{p}} u^{\intercal}\widehat{\Sigma}u \quad
&\text{subject to}&\; \|\widehat{\Sigma}u-\xnew\|_{\infty}\leq  \|\xnew\|_2 \lambda_{n} \label{eq: constraint 1 Poisson} \\
&\; &|\xnew^{\intercal}\widehat{\Sigma}u-\|\xnew\|_2^2 |\leq \|\xnew\|_2^2\lambda_{n} \label{eq: constraint 2 Poisson} \\
&\; &\|X u\|_{\infty} \leq \|\xnew\|_2 \eta_n,
\label{eq: constraint 3 Poisson}    
\end{eqnarray}
with $\lambda_{n}\asymp \left({\log p}/{n}\right)^{1/2}$ and $\eta_n \asymp \sqrt{\log n}$. Here we explain the construction of $\widehat{u}$. The variance of the first term in the error decomposition \eqref{eq: error decomposition} is of the same order of magnitude as the objective function $u^{\intercal}\widehat{\Sigma}u$ scaled by $1/n$. The first constraint \eqref{eq: constraint 1 Poisson} controls the second term, while the last constraint \eqref{eq: constraint 3 Poisson} bounds the remaining terms in \eqref{eq: error decomposition}. Consequently, the objective function, along with the first and third constraints, ensures that the error $\widehat{x_*^{\intercal}\beta}-x_*^{\intercal}\beta$ remains small. Although this strategy of minimizing the variance while constraining the bias is effective when $x_*$ has certain special structures \citep{javanmard2014confidence, zhang2014confidence, athey2018approximate, cai2017confidence}, it can be shown that this approach fails under certain dense structures of $x_*$. In such cases, the minimizer $\widehat{u}$ turns out to be zero, as demonstrated in Proposition 2 of \cite{cai2019individualized}. To address this issue, we introduce an additional constraint \eqref{eq: constraint 2 Poisson}. This constraint ensures that the variance level of $\widehat{u}^{\intercal}\frac{1}{n}\sum_{i=1}^{n}e^{-X_{i\cdot}^{\intercal}\widehat{\beta}}X_{i\cdot}\epsilon_i$ dominates the bias terms in \eqref{eq: error decomposition}, by providing a lower bound for the variance level. For more details, refer to Lemma \ref{lem: variance lower bound}.

\vspace{2 mm}
\noindent From the construction in \eqref{eq: correction}, the asymptotic variance of $\widehat{x_*^{\intercal}\beta}$ can be obtained as
\begin{equation}
    \mathrm{V}_{x_*} = \frac{1}{n^{2}}\widehat{u}^{\intercal}\sum_{i=1}^{n}e^{-X_{i\cdot}^{\intercal}\beta}X_{i\cdot}X_{i\cdot}^{\intercal}\widehat{u},
    \label{eq: asymptotic variance}
\end{equation}
which can be estimated by plugging in the estimates of the unknown parameters as
\begin{equation}
    \widehat{\mathrm{V}}_{x_*} = \frac{1}{n^{2}}\widehat{u}^{\intercal}\sum_{i=1}^{n}e^{-X_{i\cdot}^{\intercal}\widehat{\beta}}X_{i\cdot}X_{i\cdot}^{\intercal}\widehat{u}.
    \label{eq: estimated asymptotic variance}
\end{equation}
We finally construct the confidence interval for $x_*^{\intercal}\beta$ as
\begin{equation}
    \mathrm{CI}_{\alpha}(x_*) = \left[\widehat{x_*^{\intercal}\beta} - z_{\alpha/2}\widehat{\mathrm{V}}_{x_*}^{1/2}, \widehat{x_*^{\intercal}\beta} + z_{\alpha/2}\widehat{\mathrm{V}}_{x_*}^{1/2}\right],
    \label{eq: CI}
\end{equation}
where $z_{\alpha/2}$ is the upper $\alpha/2$-quantile of the standard normal distribution.  We conduct the hypothesis testing \eqref{eq: hypothesis linear}
\begin{equation}
    \phi_{x_*}(\alpha) = \mathbf{1}\left(\left|\widehat{x_*^{\intercal}\beta}\right| - z_{\alpha/2}\widehat{V}_{x_*}^{1/2} > 0\right).
    \label{eq: testing}
\end{equation}

\subsection{Inference for Quadratic Functional}
\label{sec: QF}
In this section, we propose statistical inference procedures for $Q_A := \beta_G^{\intercal}A\beta_G$ where for $G \subset \{1,2,\ldots,p\}, A \in \R^{|G|\times|G|}$ can be either a known positive definite matrix or $A = \Sigma_{G,G}$ where $\Sigma = \mathbb{E}\left(X_{i\cdot}X_{i\cdot}^{\intercal}\right)$ is the unknown population covariance matrix. For simplicity, we assume the index set $G$ is of the form $\{1,2,\ldots,|G|\}$. Define
\begin{equation*}
    \widehat{A} =
    \begin{cases}
        A \hspace{117 pt} \text{if } A \text{ is known}\\
        \widehat{\Sigma}_{G,G} = \frac{1}{n}\sum_{i=1}^{n} X_{iG}X_{iG}^{\intercal} \hspace{18 pt} \text{if } A = \Sigma_{G,G} \text{ is unknown}
    \end{cases}
\end{equation*}
As pointed out in section \ref{sec: lin quad inf} the plug-in estimator $\widehat{\beta}_G^{\intercal}\widehat{A}\widehat{\beta}_G$ needs to be bias-corrected. The error of the plug-in estimator can be decomposed as 
\begin{equation}
\widehat{\beta}_G^{\intercal}\widehat{A}\widehat{\beta}_G - \beta_G^{\intercal}A\beta_G = -2\widehat{\beta}_G^{\intercal}\widehat{A}\left(\beta_G - \widehat{\beta}_G\right) + \beta_G^{\intercal}\left(\widehat{A} - A\right)\beta_G - \left(\widehat{\beta}_G - \beta_G\right)^{\intercal}\widehat{A}\left(\widehat{\beta}_G - \beta_G\right).  
\end{equation}
Note that the dominant term $-2\widehat{\beta}_G^{\intercal}\widehat{A}\left(\beta_G - \widehat{\beta}_G\right) = -2\left(\widehat{\beta}_G^{\intercal}\widehat{A} \quad \mathbf{0}\right)\left(\beta - \widehat{\beta}\right)$ in the decomposition is a linear functional of $(\beta - \widehat{\beta})$ and hence can be estimated using the technique in section \ref{sec: LF} with $x_* := \left(\widehat{\beta}_G^{\intercal}\widehat{A}\quad \mathbf{0}\right)^{\intercal}$. Specifically, the dominant term is estimated by $\frac{2}{n}\widehat{u}_A^{\intercal}\sum_{i=1}^{n}e^{-X_{i\cdot}^{\intercal}\widehat{\beta}}X_{i\cdot}\left(y_i - e^{X_{i\cdot}^{\intercal}\widehat{\beta}}\right)$ where $\widehat{u}_A$ is the projection direction constructed in the following equation \eqref{eq: projection S}.  Note that the construction of $\widehat{u}_A$ follows from (\eqref{eq: constraint 1 Poisson}, \eqref{eq: constraint 2 Poisson}, \eqref{eq: constraint 3 Poisson}) with $x_* := \left(\widehat{\beta}_G^{\intercal}\widehat{A}\quad \mathbf{0}\right)^{\intercal}$. Group inference in the high-dimensional linear regression model is studied in detail in \cite{grouplin}. 

\noindent Finally, we propose the following bias-corrected estimator for $Q_A$
    \begin{equation}
        \widehat{Q}_A = \widehat{\beta}_G^{\intercal}\widehat{A}\widehat{\beta}_G + \frac{2}{n}\widehat{u}_A^{\intercal}\sum_{i=1}^{n}e^{-X_{i\cdot}^{\intercal}\widehat{\beta}}X_{i\cdot}\left(y_i - e^{X_{i\cdot}^{\intercal}\widehat{\beta}}\right).
        \label{eq: estimate QF}
    \end{equation}
    We construct the projection direction $\widehat{u}_A$ as \begin{equation}
        \begin{aligned}
        \widehat{u}_A=\arg \min u^{\intercal} \widehat{\Sigma} u \quad
        \text { s.t. } \max _{w \in \mathcal{C}}\left\langle w, \widehat{\Sigma} u-\left(\widehat{\beta}_G^{\intercal} \widehat{A} \quad \mathbf{0}\right)^{\intercal}\right\rangle & \leq\left\|\widehat{A} \widehat{\beta}_G\right\|_2 \lambda_n \\
        \|Xu\|_{\infty} & \leq \eta_n\|\widehat{A}\widehat{\beta}_G\|_2,
        \end{aligned}
        \label{eq: projection S}
    \end{equation}
    where $\mathcal{C}=\left\{e_1, \cdots, e_p, \frac{1}{\left\|\widehat{A} \widehat{\beta}_G\right\|_2}\left(\widehat{\beta}_G^{\intercal} \widehat{A} \quad \mathbf{0}\right)^{\intercal}\right\},\lambda_n \asymp \sqrt{\log p / n}$ and $\eta_n \asymp \sqrt{\log n}$.  
    We then estimate the variance of the proposed estimator $\widehat{Q}_A$ by
    \small{\begin{equation}
        \widehat{\mathrm{V}}_A(\tau) =
        \begin{cases}
            \frac{4}{n^2} \widehat{u}_A^{\intercal} \sum_{i=1}^{n}e^{-X_{i\cdot}^{\intercal}\widehat{\beta}}X_{i\cdot}X_{i\cdot}^{\intercal} \widehat{u}_A + \frac{\tau}{n}; \hspace{108 pt} \text{if } A \text{ is known}\\
            \\
            \frac{4}{n^2} \widehat{u}_A^{\intercal} \sum_{i=1}^{n}e^{-X_{i\cdot}^{\intercal}\widehat{\beta}}X_{i\cdot}X_{i\cdot}^{\intercal} \widehat{u}_A \hspace{4 pt} \\
            \hspace*{10 pt} + \frac{1}{n^2} \sum_{i=1}^n\left(\widehat{\beta}_G^{\intercal} X_{i G} X_{i G}^{\intercal} \widehat{\beta}_G-\widehat{\beta}_G^{\intercal} \widehat{\Sigma}_{G, G} \widehat{\beta}_G\right)^2+\frac{\tau}{n}; \hspace{40 pt} \text{if } A = \Sigma_{G,G} \text{ is unknown,}
        \end{cases}
        \label{eq: estimated enlarged var}
    \end{equation}}
    where $\tau > 0$. The constant $\tau$ is crucial for addressing super-efficiency issues, as detailed in \cite{grouplin}. Near the null hypothesis $Q_A = 0$, the variance decreases significantly faster than the typical $1/\sqrt{n}$ rate, which is known as "super-efficiency." In this scenario, the upper bound for the bias can become larger than the standard deviation, even if $k = |\beta|_0 \leq c\frac{\sqrt{n}}{\log p \log n}$. To mitigate the effects of ``super-efficiency", we increase the variance by adding $\tau /n$. This adjustment ensures that the variance consistently exceeds the upper bound for the bias. \cite{grouplin} recommends using $\tau = 0.5$ or $\tau = 1$.
    
    \vspace{2 mm}
    \noindent Naturally then we can construct a $100(1-\alpha)\%$ confidence interval for $Q_A$ as
    \begin{equation}
        \mathrm{CI}_{\alpha}(\tau) = \left[\widehat{Q}_A - z_{\alpha/2}\widehat{\mathrm{V}}_A^{1/2}(\tau), \widehat{Q}_A + z_{\alpha/2}\widehat{\mathrm{V}}_A^{1/2}(\tau)\right].
        \label{eq: CI QF}
    \end{equation}
    We propose the following $\alpha-$level significance test for testing \eqref{eq: significance}
    \begin{equation}
        \phi_A(\tau) = \mathbf{1}\left(\widehat{Q}_A - z_{\alpha}\widehat{V}_A(\tau)^{1/2} > 0\right).
        \label{eq: testing QF}
    \end{equation}
    
\section{Theoretical Properties}
\label{sec: theory}
We first introduce the following assumptions to facilitate theoretical analysis.
\begin{enumerate}
    \item[(C1)] The rows $\{X_{i\cdot}\}_{i=1}^n$ are i.i.d. $p$-dimensional {sub-Gaussian} random vectors with $\Sigma=\mathbb{E} (X_{i\cdot} X_{i\cdot}^{\intercal})$ where $\Sigma$ satisfies $c_0\leq \lambda_{\min}\left(\Sigma\right) \leq \lambda_{\max}\left(\Sigma\right) \leq C_0$ for some positive constants $C_0\geq c_0>0$ and $\max_{1\leq i \leq n}|X_{i\cdot}^{\intercal}\beta| \leq C$ for some positive constant $C>0$; The high-dimensional vector $\beta$ is assumed to be of sparsity $k$. Also assume $\exists \widetilde{C} > 0$ and $\nu > 0$ such that $\mathbb{E}\left[|\epsilon_i|^{2+\nu}\mid X_{i\cdot}\right] \leq \widetilde{C}$.
\end{enumerate}
Assumption {\rm (C1)} first imposes a tail condition on the high-dimensional covariates $X_{i\cdot}$. Next, the bounded spectrum assumption on the population covariance matrix $\Sigma$ is generally assumed in theoretical analysis of high-dimensional regression models \citep{athey2018approximate,ma2018global,GLM}. Specifically under the sub-Gaussian condition on the covariates $X$, the bounded spectrum assumption implies the restricted eigenvalue condition 
\citep{bickel2009simultaneous,zhou2009restricted,negahban2009unified,raskutti2010restricted}. We further assume $\forall i, X_{i\cdot}^{\intercal}\beta$ is bounded. Although this is a restrictive condition, it seems necessary for the theoretical analysis. In Poisson regression, the conditional mean of the outcome is an exponent of the linear predictor $X_{i\cdot}^{\intercal}\beta$. The conditional mean increases much faster than the linear predictor, and will easily explode even if the linear predictor is moderately large. A similar condition exists in the literature of high-dimensional generalized linear regression. See \cite{measerror}; assumption {\rm (A2)} in \cite{caseprob}; assumption {\rm (A4)} in \cite{poisonline}; assumption {\rm (A2)} in \cite{streaming}; and \cite{ning2017general} for Poisson regression. Additionally, we have a moment bound assumption on the random errors $\epsilon_i$'s. Such a condition has been previously used in analyzing high-dimensional linear regression; see Section 6 in \cite{javanmard2017flexible} and Proposition 1 in \cite{cai2019individualized}. 

\vspace{2 mm}
\noindent The following proposition states the theoretical properties of  the penalized  estimator $\widehat{\beta}$ in \eqref{eq: lasso}

\begin{proposition}
Suppose that Condition {\rm (C1)} holds. Then, if we choose $\lambda_0 := \left\|\frac{1}{n}\sum_{i=1}^{n}\epsilon_iX_{i\cdot}\right\|_{\infty} \asymp \sqrt{\frac{\log p}{n}}$ and assume that $\max_{i,j}\left|X_{ij}\right|k
\lambda_0 \leq c$ for some constant $c>0$, then with probability greater than $1-p^{-c}-g(n)$ the initial estimator $\widehat{\beta}$ in \eqref{eq: lasso} with $\lambda = (1+\delta_0)\lambda_0$ for any $\delta_0>0$ satisfies the following
\begin{equation}
\|\widehat{\beta}-\beta\|_1\leq C k \sqrt{\frac{\log p}{n}} \quad \text{and}\quad \|\widehat{\beta}_{S^{c}}-\beta_{S^{c}}\|_1 \leq \left(2/\delta_0 + 1\right) \|\widehat{\beta}_{S}-\beta_{S}\|_1,
 \label{eq: est property}
\end{equation}
where $S$ denotes the support of $\beta$ and $C>0$ is a positive constant.
\label{prop: lasso convergence}
\end{proposition}
\noindent The asymptotic normality established in the next subsections will hold for any initial estimator satisfying \eqref{eq: est property}, including our initial estimator defined in \eqref{eq: lasso}.

\subsection{Guarantees for Inference for \texorpdfstring{$x_*^{\intercal}\beta$}{TEXT}}
\label{sec: Inf linear}
A key step in establishing asymptotic normality of the estimator $\widehat{x_*^{\intercal}\beta}$ is to guarantee that the variance component of the estimator $\widehat{x_*^{\intercal}\beta}$ dominates its bias components. The constraint \eqref{eq: constraint 2 Poisson} plays a crucial role in guaranteeing such a dominance by providing a lower bound for the asymptotic variance.
\begin{lemma}
With probability greater than $1-p^{-c}-\exp (-c n)$,
\begin{equation}
c_1\frac{\|x_*\|_2}{\sqrt{n}} \leq \sqrt{\mathrm{V}_{x_*}} \leq C_1\frac{\|x_*\|_2}{\sqrt{n}},    
\label{eq: variance lower bound}
\end{equation}
for some positive constants $c, c_1, C_1>0$.
Here ${\mathrm V}_{x_*}$ is defined in \eqref{eq: asymptotic variance}.
\label{lem: variance lower bound}
\end{lemma}
\noindent Next, we establish the limiting distribution for the proposed point estimator $\widehat{x_*^{\intercal}\beta}$.
\begin{theorem}
Suppose that Condition {\textrm (C1)} holds and  $\eta_{n} \asymp \sqrt{\log n}$ defined in \eqref{eq: constraint 3 Poisson} satisfies $\eta_n \frac{k \log p \log n}{\sqrt{n}} \to 0$. Then for any $\widehat{\beta}$ satisfying \eqref{eq: est property}, the estimator $\widehat{x_*^{\intercal}\beta}$ defined in \eqref{eq: correction} satisfies
\begin{equation}
\frac{1}{\left(\mathrm V_{x_*}\right)^{1/2}}\left(\widehat{x_*^{\intercal}\beta}-x_*^{\intercal}\beta\right) \longrightarrow_d N(0,1),
\label{eq: limiting distribution of linear functional}
\end{equation}
where ${\mathrm V}_{x_*}$ is defined in \eqref{eq: asymptotic variance}.
\label{thm: limiting distribution}    
\end{theorem}
\noindent It is to be noted that the asymptotic normality in Theorem \ref{thm: limiting distribution} is established for arbitrary $x_*$ and no sparsity condition is imposed on the precision or the inverse Hessian matrix. This has considerably weakened the existing sparsity assumptions for the inverse Hessian matrix \citep{van2014asymptotically, ning2017general,ma2018global}. Additionally, sparsity condition similar to that in Theorem \ref{thm: limiting distribution}, $k \ll \frac{\sqrt{n}}{\log p \log n}$, has been shown to be necessary and sufficient for confidence interval construction of individual regression coefficients in high-dimensional linear and logistic regression \citep{cai2017confidence, cai2021logistic}.
\noindent Theorem \ref{thm: limiting distribution} also justifies the confidence interval construction in \eqref{eq: CI}. 

\vspace{2 mm}
\noindent To study the testing procedure, we introduce the following parameter space for $\theta=(\beta, \Sigma)$, 
$$\Theta(k)=\left\{\theta=(\beta, \Sigma): \|\beta\|_0\leq k, c_0\leq \lambda_{\min}(\Sigma)\leq \lambda_{\max}(\Sigma)\leq C_0\right\},$$ for some positive constants $C_0\geq c_0>0$. We consider the null space
$
\mathcal{H}_0=\left\{\theta=(\beta,\Sigma)\in \Theta(k): x_*^{\intercal}\beta = 0\right\}
$
and the local alternative parameter space 
\begin{equation*}
\mathcal{H}_1(t)=\left\{\theta=(\beta,\Sigma)\in \Theta(k): x_*^{\intercal}\beta =t\|x_*\|_2/{n^{1/2}}\right\}, \quad \text{for some}\; t\neq 0.
\label{eq: alter space}
\end{equation*}

\begin{proposition}
Under the same conditions as in Theorem \ref{thm: limiting distribution}, for each $\theta\in \mathcal{H}_0$, the proposed testing procedure ${\phi}_{x_*}(\alpha)$ in \eqref{eq: testing} satisfies
$
\lim_{n\to \infty}\mathbb{P}_{\theta}\left[{\phi}_{x_*}(\alpha)=1\right] = \alpha.
$ 
For $\theta\in \mathcal{H}_1(t)$, we have 
\begin{equation}
\lim_{n \longrightarrow \infty}\mathbb{P}_{\theta}\left[\phi_{x_*}(\alpha) = 1\right] = 1 + \Phi\left(-z_{\alpha/2} + \frac{t\|x_*\|_2/\sqrt{n}}{\sqrt{\rm V}_{x_*}}\right) - \Phi\left(z_{\alpha/2} + \frac{t\|x_*\|_2/\sqrt{n}}{\sqrt{\rm V}_{x_*}}\right),
\label{eq: power}
\end{equation}
where $\Phi$ is the cumulative distribution function of standard normal distribution.
\label{prop: testing}    
\end{proposition}
\noindent The proposed hypothesis testing procedure is shown to have a well-controlled type I error rate. Since \eqref{eq: variance lower bound} implies $\left(n {\rm V}_{x_*}\right)^{1/2} \asymp \|\xnew\|_2$, therefore if $t \to \infty$, the power will be 1 in the asymptotic sense.

\subsection{Guarantees for Inference for \texorpdfstring{$Q_A$}{TEXT}}
\label{sec: Inf QF}
Here we introduce an additional assumption to facilitate the theoretical proofs.
\begin{enumerate}
    \item[(C2)] $\widehat{\beta}$ is independent of the data $(X, y)$ used in the construction of \eqref{eq: estimate QF}. 
\end{enumerate}
Such independence can be achieved by employing a sample-splitting approach. Here, we begin by randomly dividing the data into two distinct subsamples, denoted as $\left(X^{(1)},y^{(1)}\right)$ with a sample size of $n_1$, and $\left(X^{(2)},y^{(2)}\right)$ with a sample size of $n_2 = n - n_1$. We then proceed to construct $\widehat{\beta}$ in \eqref{eq: lasso} using the $\left(X^{(1)},y^{(1)}\right)$ subsample, followed by implementing the bias correction procedure in \eqref{eq: estimate QF} based on the data from the $\left(X^{(2)},y^{(2)}\right)$ subsample. We believe the condition is not essential for the proposed method and is introduced for technical convenience. In fact the simulation study demonstrates that the proposed method performs well numerically without splitting, that is even when the independence assumption is not met. Next, we characterize the estimator $\widehat{Q}_A$.

\begin{theorem}
    Suppose that Condition {\textrm (C1)} holds, 
    then for any estimator $\widehat{\beta}$ satisfying {\eqref{eq: est property}} and {\textrm (C2)}, as $n,p \to \infty$, $\widehat{Q}_A - Q_A = M_A + B_A$ where
    \begin{eqnarray}
        \frac{M_A}{\sqrt{V_A}} &\to & N(0,1) \label{eq: asymptotic normal QF} \\
        \mathbb{P}\left(|B_A|  \lesssim \left\|\widehat{A}\widehat{\beta}_G\right\|_2 \frac{k \log p \log n}{n} + \|A\|_2\frac{k \log p}{n}\right) & \geq & 1 - p^{-c} - g(n) - e^{-\sqrt{n}},
    \label{eq: limiting distribution QF}
    \end{eqnarray}
    with \small{\begin{equation*}
        \mathrm{V}_A =
        \begin{cases}
            \frac{4}{n^2} \widehat{u}_A^{\intercal} \sum_{i=1}^{n}e^{-X_{i\cdot}^{\intercal}\beta}X_{i\cdot}X_{i\cdot}^{\intercal} \widehat{u}_A ; \hspace{170 pt} \text{if } A \text{ is known}\\
            \frac{4}{n^2} \widehat{u}_A^{\intercal} \sum_{i=1}^{n}e^{-X_{i\cdot}^{\intercal}\beta}X_{i\cdot}X_{i\cdot}^{\intercal} \widehat{u}_A + \frac{1}{n} \mathbb{E}\left(\beta_G^{\intercal} X_{i G} X_{i G}^{\intercal} \beta_G-\beta_G^{\intercal} \Sigma_{G, G} \beta_G\right)^2 ; \quad \quad \text{if } A = \Sigma_{G,G} \text{ is unknown.}
        \end{cases}
    \end{equation*}}
\label{thm: limiting distribution QF}
\end{theorem}
\noindent The theorem above demonstrates that the principal error component $M_A$ converges to an asymptotic normal distribution. The variance level differs with choice of $A$, when $A = \Sigma_{G,G}$ the variance consists of the uncertainty of estimating both $\beta$ and $\Sigma$ while when $A$ is known the variance has to only take into account the estimation of $\beta$. 

\vspace{2 mm}

\noindent Theorem \ref{thm: limiting distribution QF} justifies the confidence interval construction in \eqref{eq: CI QF}. To study the testing procedure \eqref{eq: testing QF}, consider the following null parameter space
$
\widetilde{\mathcal{H}}_0=\left\{\theta=(\beta,\Sigma)\in \Theta(k): \|\beta_G\|_2 = 0\right\}
$
and the local alternative parameter space, for $t > 0$, 
\begin{equation*}
\widetilde{\mathcal{H}}_{1, A}(t)=\left\{(\beta, \Sigma) \in \Theta(k): \beta_G^{\intercal} A \beta_G=(1.01z_{\alpha}+t)\sqrt{\rm V_A}\right\}.
\label{eq: alter space Q}
\end{equation*}
\vspace{-5mm}
\begin{proposition}
Suppose that Condition {\textrm (C1)} holds, $\eta_{n} \asymp \sqrt{\log n}$ defined in \eqref{eq: projection S} satisfies

$\eta_n \frac{k \log p \log n}{\sqrt{n}}\to 0$. Then for any estimator $\widehat{\beta}$ satisfying {\eqref{eq: est property}} and {\textrm (C2)}, for each $\theta\in \widetilde{\mathcal{H}}_0$, the proposed testing procedure ${\phi}_A(\tau)$ in \eqref{eq: testing QF} satisfies
$
\lim_{n\to \infty}\mathbb{P}_{\theta}\left[{\phi}_A(\tau)=1\right] = \alpha.
$
For $\theta\in \widetilde{\mathcal{H}}_{1, A}(t)$ 

\begin{equation}
\liminf_{n \longrightarrow \infty}\mathbb{P}_{\theta}\left[\phi_A(\tau) = 1\right] \geq 1 - \Phi\left( - t\right).
\label{eq: power QF}
\end{equation}
\label{prop: testing QF}
\end{proposition}
\noindent The proposed hypothesis testing procedure \eqref{eq: testing QF} is shown to have a well-controlled type I error rate and has power converging to $1$ as long as $t \to \infty$.

\section{Application to Mediation Analysis}
\label{sec: mediation}

Mediation analysis is a statistical technique used to investigate the mechanisms underlying the relationship between an independent variable and a dependent variable, mediated through intervening variables. Regression based approaches are widely used in mediation analysis to estimate the direct and indirect effects of an independent variable on a dependent variable \citep{regmed1, regmed2}. As discussed in section \ref{sec: intro}, in mediation analysis we often encounter with high-dimensional data, examples include genomics, neuro-imaging and social sciences  \citep{medapp1, medapp2, medapp3, medapp4, medapp5}. High-dimensional mediation analysis with continuous outcome is studied in \cite{hdmed, PTSD, mednus}. \cite{hdmed} studied the natural indirect effect of the high-dimensional mediators on a continuous outcome using a debiased lasso estimator under the assumption of sparse $\gamma$ in \eqref{eq: mediation model}. \cite{mednus} further relaxed such sparsity assumptions and also addressed the challenging situation where both exposure–mediator and mediator–outcome coefficients are zero. \cite{PTSD} used desparsified Lasso estimates to conclude about significant mediators, however, without stating any theoretical guarantee. We apply the debiasing techniques in section \ref{sec: Inf linear} to solve the case with count outcome following a Poisson distribution while allowing for arbitrary dense $\gamma$. In our study, we focus on a simplified case of mediation analysis. Unlike \cite{medinter1, medinter2, medinter3} that address more intricate mediation analysis problems, we specifically do not take into account the interaction factor among the high-dimensional mediators. By intentionally excluding this aspect, we aim to provide a clearer understanding of the basic mechanisms involved in the mediation process when the outcome is a count variable.

\vspace{2 mm}
\noindent For the $i$ th subject, $i=1, \ldots, n$, we use $y_i$ to denote the count outcome, $M_i$ to denote a vector of $p$ mediators,  and $T_i$ to denote the treatment. We consider the high-dimensional regime where the number of mediators $p$ is allowed to be much larger than the sample size $n$. We consider the following mediation model:
    \begin{equation}
        y_i \sim \text{Poisson}\left(e^{M_i^{\intercal} \beta_0+T_i\beta_1 + T_i.M_i^{\intercal}\beta_2}\right), \quad M_i = \gamma T_i + E_i, \quad 1\leq i \leq n,        
       \label{eq: mediation model}
    \end{equation}
    where $E_i$ are mean-zero random vectors that are uncorrelated with $T_i$. Define 
    \begin{equation}
        X_{i\cdot} := \left(M_i^{\intercal}, T_i, T_i\cdot M_i^{\intercal}\right)^{\intercal} \; \text{and } \; \beta := \left(\beta_0^{\intercal}, \beta_1, \beta_2^{\intercal}\right)^{\intercal}.
        \label{eq: define X and b}
    \end{equation}
    Then model \eqref{eq: mediation model} can be expressed as the familiar Poisson regression model, $y_i \sim \text{Poisson}\left(e^{X_{i\cdot}^{\intercal}\beta}\right)$ for $1 \leq i \leq n$.

    \vspace{2 mm}
\noindent \underline{\textbf{Interaction Test}}: 
Consider the interaction term $\beta_2$ in \eqref{eq: mediation model}. With \eqref{eq: define X and b}, the test of interaction $H_0: \beta_2 = 0$ can be reformulated as $H_0: \beta_G = 0$ with $G = \{p+2,\ldots,2p+1\}$. We can then apply the procedures developed in section \ref{sec: QF} to test the existence of interaction. One possible way is to perform the test in \eqref{eq: testing QF} with $A = \Sigma_{G,G}$ where $\Sigma = \mathbb{E}X_{i\cdot}X_{i\cdot}^{\intercal}$. Precisely, for a fixed $\tau > 0$, we propose the following test to test the hypothesis $H_0 : \beta_2 = 0$ of no significant interaction between the exposure and mediator
\begin{equation}
    {\phi}_{\beta_2}(\tau) = \mathbf{1}\left\{\widehat{\beta_G^{\intercal}\Sigma_{G,G}\beta_G} > z_{\alpha}\widehat{V}_{\Sigma}^{1/2}(\tau)\right\},
    \label{eq: testing interaction}
\end{equation}
where $\widehat{\beta_G^{\intercal}\Sigma_{G,G}\beta_G}$ and $\widehat{V}_{\Sigma}(\tau)$ are given by \eqref{eq: estimate QF} and \eqref{eq: estimated enlarged var}  with $A = \Sigma_{G,G}$, $\widehat{A} = \widehat{\Sigma}_{G,G}$ and $G = \{p+2,\ldots,2p+1\}$. The type I error control and the power results for ${\phi}_{\beta_2}(\tau)$ are summarized in the following proposition.

\begin{proposition}
    Consider the definitions in \eqref{eq: define X and b}. Suppose Condition {\rm (C1)} holds and $\eta_{n}$ defined in \eqref{eq: projection S} satisfies $\eta_n \frac{k \log p \log n}{\sqrt{n}} \to 0$. Then for any estimator $\widehat{\beta}$ satisfying assumption {\eqref{eq: est property}} and {\textrm (C2)}, for each $\theta\in \widetilde{\mathcal{H}}_0$, the proposed testing procedure ${\phi}_{\beta_2}(\tau)$ in \eqref{eq: testing interaction} satisfies
    $
    \lim_{n\to \infty}\mathbb{P}_{\theta}\left[{\phi}_{\beta_2}(\tau)=1\right] = \alpha.
    $
    For $\theta\in \widetilde{\mathcal{H}}_{1, \Sigma_{G,G}}(t)$  
    \begin{equation}
    \liminf_{n \longrightarrow \infty}\mathbb{P}_{\theta}\left[\phi_{\beta_2}(\tau) = 1\right] \geq 1 - \Phi\left( - t\right),
    \label{eq: power interaction}
    \end{equation}
    $\widetilde{\mathcal{H}}_0$ and $\widetilde{\mathcal{H}}_{1,\Sigma_{G,G}}(t)$ are defined in section \ref{sec: Inf QF} with $A = \Sigma_{G,G}$. 
    \label{prop: testing interaction}
    \end{proposition}

\noindent {\underline{\bf Test of Indirect Effect in Absence of Interaction }}: 
Consider the mediation model \eqref{eq: mediation model} with no interaction term i.e., $\beta_2 = 0$.
\begin{equation}
        y_i \sim \text{Poisson}\left(e^{M_i^{\intercal} \beta_0+T_i\beta_1}\right), \quad M_i = \gamma T_i + E_i, \quad 1\leq i \leq n.        
       \label{eq: mediation model wo int}
\end{equation}
Note that model \eqref{eq: mediation model wo int} implies that $M_i^{\intercal} \beta_0=T_i \gamma^{\intercal} \beta_0+\epsilon_{2 i}$, where $\epsilon_{2 i}=E_i^{\intercal} \beta_0$.  Then the quantity $\gamma^{\intercal} \beta_0$ can be viewed as the indirect effect of the exposure on the count response, mediated through $M_i$. With \eqref{eq: define X and b}, $\gamma^{\intercal}\beta_0$ can be re-defined as a linear functional of $\beta = (\beta_0^{\intercal} \quad \beta_1)^{\intercal}$ explicitly as $\left(\gamma^{\intercal} \quad 0\right)\beta$. However, the loading $\left(\gamma^{\intercal} \quad 0\right)^{\intercal}$ involves unknown parameter $\gamma$. So for $j = 1,2,\ldots,p$ we estimate $\gamma_j$ by the following OLS estimator;
\begin{equation}
    \widehat{\gamma}_j = \arg\min_{\gamma_j \in \mathbb{R}} \frac{1}{2n}\sum_{i=1}^{n}\left(M_{ij} - \gamma_jT_i\right)^{2} = \frac{\sum_{i=1}^{n}M_{ij}T_i}{\sum_{i=1}^{n}T_i^{2}}.
    \label{eq: est gamma}
\end{equation}
To conduct inference for $\gamma^{\intercal} \beta_0$, we then apply the analysis in section \ref{sec: LF} with $x_* := (\widehat{\gamma}^{\intercal} \quad 0)^{\intercal}$ estimating $\left(\gamma^{\intercal} \quad 0\right)^{\intercal}$. Specifically, the bias-corrected estimator for $\gamma^{\intercal}\beta_0$ is given by
\begin{equation}
    \widehat{\gamma^{\intercal}\beta_0} = \widehat{\gamma}^{\intercal}\widehat{\beta}_0 + \frac{1}{n}\widehat{u}_{\gamma}^{\intercal}\sum_{i=1}^{n}e^{-X_{i\cdot}^{\intercal}\widehat{\beta}}X_{i\cdot}\left(y_i - e^{X_{i}^{\intercal}\widehat{\beta}}\right),
    \label{eq: debias}
\end{equation}
where $\widehat{u}_{\gamma}$ is constructed as (\eqref{eq: constraint 1 Poisson}, \eqref{eq: constraint 2 Poisson}, \eqref{eq: constraint 3 Poisson}) with $x_* := (\widehat{\gamma}^{\intercal} \quad 0)^{\intercal}$. We reject the hypothesis $H_0 : \gamma^{\intercal}\beta_0 = 0$ of no significant indirect effect of the exposure variable if 
\begin{equation}
\left|\widehat{\gamma^{\intercal}\beta_0}\right| > z_{\alpha/2}\widehat{\mathrm V}_{\gamma},
\label{eq: testing med}
\end{equation}
where $\widehat{\mathrm V}_{\gamma} := \frac{1}{n^2}\widehat{u}_{\gamma}^{\intercal}\sum_{i=1}^{n}e^{-X_{i\cdot}^{\intercal}\widehat{\beta}}X_{i\cdot}X_{i\cdot}^{\intercal}\widehat{u}_{\gamma} + \frac{\widehat{\sigma}_2^{2}}{\sum_{i=1}^{n}T_i^{2}}$ and $\sigma_2^2 = \operatorname{Var}(\epsilon_{2i})$ is estimated as 

\noindent $\widehat{\sigma}_2^2 := \widehat{\beta}_0^{\intercal}\frac{1}{n}\sum_{i=1}^{n}\left(M_i - \widehat{\gamma}T_i\right)\left(M_i - \widehat{\gamma}T_i\right)^{\intercal}\widehat{\beta}_0$. The additional term $\frac{\widehat{\sigma}2^{2}}{\sum{i=1}^{n}T_i^{2}}$ accounts for the uncertainty introduced by estimating $\gamma$ with $\widehat{\gamma}$, as described in \eqref{eq: est gamma}. The following theorem justifies the proposed testing procedure \eqref{eq: testing med}.

\begin{theorem}
Consider the definitions in \eqref{eq: define X and b}. Suppose Condition {\rm (C1)} holds and $\eta_{n}$ defined in \eqref{eq: constraint 3 Poisson} satisfies $\eta_n \frac{k \log p \log n}{\sqrt{n}} \to 0$. Then for any estimator $\widehat{\beta}$ satisfying assumption {\eqref{eq: est property}}, 
\begin{equation}
\sqrt{n}\left(\widehat{\gamma^{\intercal}\beta_0} - \gamma^{\intercal}\beta_0\right) \longrightarrow N\left(0, \frac{1}{n}\widehat{u}_{\gamma}^{\intercal}\sum_{i=1}^{n}e^{-X_{i\cdot}^{\intercal}\beta}X_{i\cdot}X_{i\cdot}^{\intercal}\widehat{u}_{\gamma} + \sigma_2^{2}\mathbb{E}\left(\frac{1}{\frac{1}{n}\sum_{i=1}^{n}T_i^{2}}\right)\right).
\label{eq: limiting distribution of indirect effect}
\end{equation}
\label{thm: limiting distribution mediation}
\end{theorem}

\section{Numerical Studies}
\label{sec: sim}
We first discuss how to implement the estimators defined in \eqref{eq: correction} and \eqref{eq: estimate QF}. The initial estimator $\widehat{\beta}$ in \eqref{eq: lasso} is computed using the R-package \texttt{glmnet} \citep{glmnet} with the tuning parameter $\lambda$ chosen by cross-validation. The construction of projection directions in (\eqref{eq: constraint 1 Poisson}, \eqref{eq: constraint 2 Poisson}, \eqref{eq: constraint 3 Poisson}) and (\ref{eq: projection S}) are key to the bias correction step. In the following, we introduce the equivalent dual problem of constructing the projection directions. 
The constrained optimizer $\widehat{u} \in \mathbb{R}^{p}$ in (\eqref{eq: constraint 1 Poisson}, \eqref{eq: constraint 2 Poisson}, \eqref{eq: constraint 3 Poisson}) can be computed in the form of $\widehat{u}=-\frac{1}{2}\left[\widehat{v}_{-1}+\frac{x_*}{\left\|x_*\right\|_{2}} \widehat{v}_{1}\right],$ 
where $\widehat{v} \in \mathbb{R}^{p+1}$ is defined as
\begin{equation}
\widehat{v}=\underset{v \in \mathbb{R}^{p+1}}{\arg \min }\left\{\frac{1}{4} v^{\intercal} \mathbb{H}^{\intercal} \widehat{\Sigma} \mathbb{H} v+x_*^{\intercal} \mathbb{H} v+\lambda_n\left\|x_*\right\|_{2} \cdot\|v\|_{1}\right\},
\label{eq: projection dual}
\end{equation}
with $\mathbb{H}=\left[\frac{x_*}{\left\|x_*\right\|_{2}}, \mathbb{I}_{p \times p}\right] \in \mathbb{R}^{p \times(p+1)}$. $\widehat{u}_A$ in \eqref{eq: projection S} is constructed similarly with $x_* = \left(\widehat{\beta}_{G}^{\intercal} \widehat{A}_{G,G} \quad \mathbf{0}\right)^{\intercal}$. We refer to Proposition $2$ in \citet{cai2019individualized} for the detailed derivation of the dual problem \eqref{eq: projection dual}. When the tuning parameter $\lambda_n$ approaches 0 and $\widehat{\Sigma}$ is singular in the dual problem, the minimum value \eqref{eq: projection dual} tends to negative infinity. To address this issue, we choose the smallest value of $\lambda_n$ that ensures the dual problem has a finite minimum. Such selection of the tuning parameter dates at least back to \citet{javanmard2014confidence}.

\subsection{Linear Functional}
We compare our proposed estimator with the plug-in \texttt{SSHDI} estimator. The \texttt{SSHDI} package available in \url{https://github.com/feizhe/SSHDI} computes the Splitting and Smoothing for Generalized Linear Model (SSGLM) estimator $\widehat{\beta}_{\texttt{SSHDI}}$ by implementing Algorithm 1 of \cite{sshdiglm}. The estimator is computed in two steps. To obtain the maximum likelihood estimate of $\beta_j$ for $j = 1,\ldots,p$, the first step is to conduct model selection and the second step is to fit a low-dimensional GLM by regressing the response on the covariates selected from the first step along with $X_{j}$. The variable selection and the inference are conducted on separate random samples. To overcome sensitivity with respect to the choice of splitting the entire sample, the sample-splitting method is run a large number of times 
and the estimator is aggregated over such replications. The plug-in estimator of $x_*^{\intercal}\beta$ and the associated model-free variance estimator are given by $x_*^{\intercal}\widehat{\beta}_{\texttt{SSHDI}}$ and $\widehat{\mathrm V}_{\texttt{SSHDI}}$ respectively. We compare the estimators in terms of Root Mean Square Error (RMSE), standard error and bias. We also compare from the perspective of CI construction. Recall that our proposed CI is constructed as in \eqref{eq: CI}. The CI based on the plug-in \texttt{SSHDI} is constructed as,
\begin{equation*}
{\mathrm CI}_{\alpha}(\xnew)=\left[x_*^{\intercal} \widehat{\beta}_{\texttt{SSHDI}}-z_{\alpha/2}\widehat{\mathrm V}_{\texttt{SSHDI}}^{1/2}, x_*^{\intercal} \widehat{\beta}_{\texttt{SSHDI}}+z_{\alpha/2}\widehat{\mathrm V}_{\texttt{SSHDI}}^{1/2}\right].
\end{equation*}
Throughout, we let $p = 501$ including intercept and vary sample size $n$ across $\{250,350,500,800\}$. The covariates matrix $X_{-1}$ is simulated from the multivariate Gaussian distribution with zero mean and covariance $\Sigma_1 = [0.08\times 0.5^{|i-j|}]_{(p-1)\times (p-1)}$. Note that $X_{\cdot,1}$ is a column of 1's to accommodate for the intercept. We generate both exactly sparse and approximate sparse regression vector $\beta$. 
\begin{itemize}
    \item {\bf Exactly Sparse Regression :} The non-zero coordinates of the 8-sparse regression vector $\beta$ are $4,6,\ldots,18$. These non-zero entries are 8 values equally spaced in the interval $[1,2]$.

    \item {\bf Approximate Sparse Regression :} More interpretable models include small number of significant coefficients while allowing for other predictor variables to be weakly related to the outcome. To represent these practical settings, we generate $\beta$ as $\beta_1 = 0.2$, $\beta_j \sim N(0,1)$ independently for $2 \leq j \leq 10$, $\beta_j = (j-1)^{-0.5}$ for $11 \leq j \leq p$ 
\end{itemize}
Given the high-dimensional covariates we can now generate the Poisson outcome as $y_i \sim \text{Poisson}(e^{X_{i\cdot}^{\intercal}\beta})$ for $1 \leq i \leq n$. All simulation results are averaged over 500 replications. To emphasize that our method works for arbitrary loading we simulate the following $x_*$. In the first step we set $x_{\text{basis},1} = 1$ and generate $x_{\text{basis},-1} \sim N(0,\Sigma_2)$ with $\Sigma_2 = [0.04\times 0.75^{|i-j|}]_{(p-1)\times (p-1)}$. In the second step, we generate $x_* \in \mathbb{R}^p$ as a shrunk version of $x_{\text{basis}}$.
\begin{equation}
x_{*,j}=\begin{cases} 
x_{{\mathrm basis},j} &  \text{for} \; 1\leq j\leq 11\\
r \cdot x_{{\mathrm basis},j} &  \text{for}\; 12\leq j \leq 501,\\
\end{cases}
\label{eq: shrink def}
\end{equation} 
where the ratio $r$ varies across $\{1/2,1/5,1/25\}$. Here the scale parameter $r$ controls the magnitude of the noise in $x_*$ in the sense that $\|x_*\|_2$ decreases with $r$.

\begin{table}[htp!]
\centering
\caption{``r" represents the shrinkage parameter. The columns indexed with ``Cov" and ``Len" represent the empirical coverage and length of the CIs; ``t" represents the averaged computation time (in seconds). The columns indexed with ``Bias", ``SE" and ``RMSE" represent the bias, standard error and RMSE respectively. The columns under ``Proposed" and ``\texttt{SSHDI}" correspond to the proposed estimator and the plug-in \texttt{SSHDI} estimators respectively.}
\label{tab: Ex App}
\scalebox{0.6}{
\begin{tabular}{rr|r|rrrrrr|rrrrrr}
  \hline
\multicolumn{15}{c}{{\bf Sparse Regression}} \\
  \hline
  \multicolumn{2}{c|}{}&&\multicolumn{6}{c|}{Proposed}&\multicolumn{6}{c}{SSHDI} \\
   \hline
$\|x_*\|_{2}$ &{\textrm r}&$n$& Cov & Len & t & Bias & SE & RMSE & Cov & Len & t & Bias & SE & RMSE \\
\hline
 \multirow{4}{*}{2.23} & \multirow{3}{*}{$\frac{1}{2}$} &250& 0.97 & 2.32 & 15 & -0.02 & 0.42 & 0.42 & 0.99 & 4.74 & 415 & -0.10 & 0.73 & 0.73 \\ 
 & & 400& 0.96 & 1.92 & 52 & -0.01 & 0.38 & 0.38 & 0.98 & 3.84 & 598 & -0.06 & 0.66 & 0.66 \\ 
 & & 500& 0.96 & 1.65 & 110 & 0.01 & 0.33 & 0.33 & 0.99 & 3.28 & 765 & -0.05 & 0.45 & 0.45 \\ 
 & & 800& 0.95 & 1.28 & 198 & 0.01 & 0.28 & 0.28 & 0.99 & 2.85 & 1514 & -0.06 & 0.31 & 0.31 \\
 \hline
 \multirow{4}{*}{0.98} & \multirow{3}{*}{$\frac{1}{5}$} &250& 0.97 & 1.07 & 15 & -0.03 & 0.21 & 0.21 & 0.97 & 2.08 & 420 & 0.14 & 0.34 & 0.34 \\ 
 & & 400& 0.97 & 0.89 & 52 & -0.02 & 0.19 & 0.19 & 0.98 & 1.67 & 598 & 0.10 & 0.28 & 0.28 \\ 
 & & 500& 0.96 & 0.78 & 112 & -0.01 & 0.16 & 0.16 & 0.98 & 1.43 & 765 & 0.04 & 0.18 & 0.18 \\
 & & 800 & 0.95 & 0.60 & 199 & 0.00 & 0.14 & 0.14 & 0.99 & 1.23 & 1512 & 0.02 & 0.14 & 0.14 \\
 \hline
 \multirow{4}{*}{0.48} & \multirow{3}{*}{$\frac{1}{25}$} &250& 0.95 & 0.66 & 14 & -0.04 & 0.14 & 0.14 & 0.89 & 0.94 & 420 & 0.25 & 0.17 & 0.30 \\ 
  & &400& 0.95 & 0.61 & 51 & -0.02 & 0.13 & 0.13 & 0.91 & 0.76 & 600 & 0.19 & 0.13 & 0.23 \\ 
  & &500& 0.96 & 0.52 & 110 & -0.01 & 0.12 & 0.12 & 0.92 & 0.61 & 765 & 0.11 & 0.08 & 0.14 \\ 
  & &800& 0.94 & 0.39 & 198 & -0.00 & 0.10 & 0.10 & 0.93 & 0.56 & 1513 & 0.07 & 0.07 & 0.10 \\
 \hline
  \hline
   \multicolumn{15}{c}{{\bf Approximate Sparse Regression }}  \\
   \hline
  \multicolumn{2}{c|}{}&&\multicolumn{6}{c|}{Proposed}&\multicolumn{6}{c}{SSHDI}\\
   \hline
$\|x_*\|_2$ &{\textrm r}&$n$& Cov & Len & t & Bias & SE & RMSE & Cov & Len & t & Bias & SE & RMSE \\
\hline
 \multirow{4}{*}{2.23} & \multirow{3}{*}{$\frac{1}{2}$} &250& 0.96 & 2.54 & 15 & 0.11 & 0.44 & 0.46 & 0.78 & 5.58 & 415 & -1.64 & 1.47 & 2.20 \\
 & &400& 0.95 & 2.14 & 52 & 0.09 & 0.41 & 0.42 & 0.80 & 5.28 & 597 & -1.41 & 1.51 & 2.06 \\ 
 & &500& 0.96 & 1.98 & 111 & 0.07 & 0.38 & 0.38 & 0.86 & 4.13 & 764 & -1.35 & 1.06 & 1.72 \\
 & &800& 0.96 & 1.62 & 198 & 0.04 & 0.34 & 0.34 & 0.88 & 3.80 & 1514 & -0.97 & 0.98 & 1.38 \\
  \hline
 \multirow{4}{*}{0.98} & \multirow{3}{*}{$\frac{1}{5}$} &250& 0.96 & 1.17 & 14 & 0.05 & 0.22 & 0.22 & 0.65 & 3.65 & 414 & -0.83 & 0.63 & 1.04 \\
 & &400& 0.95 & 0.99 & 52 & 0.04 & 0.20 & 0.20 & 0.78 & 3.59 & 599 & -0.78 & 0.52 & 0.94 \\ 
 & &500& 0.96 & 0.82 & 110 & 0.02 & 0.18 & 0.18 & 0.85 & 3.48 & 765 & -0.63 & 0.42 & 0.76 \\
 & &800& 0.95 & 0.75 & 199 & 0.01 & 0.16 & 0.16 & 0.86 & 3.27 & 1514 & -0.55 & 0.38 & 0.67 \\
 \hline
 \multirow{4}{*}{0.48} & \multirow{3}{*}{$\frac{1}{25}$} &250& 0.94 & 0.72 & 15 & -0.03 & 0.17 & 0.17 & 0.57 & 2.97 & 416 & -0.40 & 0.32 & 0.51 \\ 
 & &400& 0.96 & 0.67 & 51 & -0.01 & 0.15 & 0.15 & 0.65 & 1.72 & 599 & -0.32 & 0.29 & 0.43 \\ 
 & &500& 0.96 & 0.54 & 109 & -0.00 & 0.13 & 0.13 & 0.74 & 1.47 & 765 & -0.20 & 0.17 & 0.26 \\
 & &800& 0.95 & 0.48 & 198 & -0.00 & 0.11 & 0.11 & 0.84 & 1.39 & 1514 & -0.15 & 0.12 & 0.19 \\
 \hline
\end{tabular}
}
\end{table}

\noindent In Table~\ref{tab: Ex App}, we compare the proposed method with \texttt{SSHDI} in terms of CI construction, the bias, the standard error deviation and also the Root Mean Squared Error (RMSE). The CIs constructed by our method have coverage over different scenarios and the lengths are reduced when a larger sample size is used to construct the CI. 
\texttt{SSHDI} suffers from the issue of overcoverage in case of sparse regression while it undercovers in case of approximate sparse $\beta$. A key assumption for valid inference using \texttt{SSHDI} algorithm is the sure screening property i.e., the true model is selected by the variable screening procedure with probability tending to 1. Under the approximate sparse regression setting \texttt{SSHDI} fails to satisfy such a restrictive condition by ignoring the small yet non-zero regression coefficients leading to substantial omitted variable bias.

\noindent We have further reported the average computation time for each method under the column ``t" (the units are in seconds). Our proposed method produces correct CIs within 2 minutes on average, whereas for $n = 600$, it requires more than 10 minutes to implement the \texttt{SSHDI} algorithm. The main reason is that the \texttt{SSHDI} algorithm is not designed to handle linear functionals of the regression coefficients and requires to run the two step procedure across $p \times B$ replications, $B$ being the number of times the sample splitting method is run for each of the $p$ regression coefficients.

\subsection{Group Significance}

To implement the group significance test proposed in the current paper, we consider four specific tests $\phi_{I}(0), \phi_{I}(1), \phi_{\Sigma}(0), \phi_{\Sigma}(1)$ where $\phi_{I}(0), \phi_{I}(1)$ are defined in \eqref{eq: testing QF} by taking $A = I$ with $\tau = 0$ and $\tau = 1$ respectively and $\phi_{\Sigma}(0), \phi_{\Sigma}(1)$ are defined in \eqref{eq: testing QF} by taking $A = \Sigma_{G,G}$ with $\tau = 0$ and $\tau = 1$ respectively. $\tau = 0$ indicates the situation where we disregard the uncertainty due to the bias, while by setting $\tau = 1$ we provide a conservative upper bound for the bias component.

\vspace{2 mm}
\noindent We generate the covariates $\{X_{i,}\}_{1\leq i\leq n}$ from the multivariate normal distribution with zero mean and covariance matrix  $\Sigma = \{0.5^{1+|j-l|}\}_{1\leq j,l\leq 500}$. We consider three regression settings. In the settings {\rm (R2)} and {\rm (R3)} the regression vectors are relatively dense with small non-zero entries.
    \begin{enumerate}
        \item[(R1)] $\beta_j = 0$ for $1\leq j \leq p$.
    
        \item[(R2)] $\beta_1 = 0 \; ; \beta_j = (j-1)^{-1}$ for $2 \leq j \leq 26 \; ; \beta_j = 0$ otherwise.

        \item[(R3)] $\beta_1 = 0 \; ; \beta_j = \frac{j-1}{50}$ for $2 \leq j \leq 21 \; ; \beta_j = 0$ otherwise.
    \end{enumerate}
    Finally we simulate the count outcome as $y_i \sim \text{Poisson}\left(e^{X_{i\cdot}^{\intercal}\beta}\right)$ for $1 \leq i \leq n$. Here sample size $n$ varies across $\{350, 500, 800\}$. We consider testing $H_0 : \beta_G = 0$ for $G = \{15, 16, \ldots, 200\}$. Table~\ref{tab: dense group} summarizes the hypothesis testing results.
    Moreover we have tabulated the coverage properties of the confidence intervals $CI_{I}(\tau)$ for $\|\beta_G\|_2^2$ and $CI_{\Sigma}(\tau)$ for $\beta_G^{\intercal}\Sigma_{G,G}\beta_G$ when $\tau = 0$ and $\tau = 1$. The results presented are the average of 500 simulations.
    
\noindent We notice that the confidence intervals with $\tau = 1$ provide proper coverage, while those constructed with $\tau = 0$ fail to do so in settings {\rm (R1)} and {\rm (R2)}. As mentioned in section \ref{sec: QF}, the ``super-efficiency" phenomenon occurs near the null hypothesis $Q_A = 0$ whence the variance is enlarged by the level $\tau/n$ to dominate the bias. In settings {\rm (R1)} and {\rm (R2)}, where the true value is close to 0, using $\tau = 0$ leads to uncertainty from the variance component being considered, but the bias component is neglected, ultimately causing the bias to overpower the variance.

\vspace{2 mm}

\noindent We further report the empirical rejection rate, the proportion of the null hypothesis being rejected out of the 500 replications. Under the null hypothesis, it is an empirical measure of the type I error while under the alternative hypothesis, it is an empirical measure of the power. Specifically, for setting {\rm (R1)} ($\beta_G = 0$), the empirical rejection rate is an empirical measure of the type I error while for settings {\rm (R2)} and {\rm (R3)} this is an empirical measure of the power of the test. We have observed that, for $\tau = 0$, the testing procedures $\phi_{\mathrm I}(0)$ and $\phi_{\Sigma}(0)$ do not control the type I error, the reason being $\tau = 0$ does not quantify the uncertainty of the bias component. In contrast, as long as we set $\tau = 1$ thereby providing a conservative upper bound for the bias component, the proposed procedures $\phi_{\mathrm I}(1)$ and $\phi_{\Sigma}(1)$ control the type I error. To compare the power, we focus on $\phi_{\mathrm I}(1)$ and $\phi_{\Sigma}(1)$ and observe that $\phi_{\mathrm{I}}(1)$ is the best. An interesting observation is that, although the proposed testing procedures $\phi_{\mathrm I}(1)$ and $\phi_{\Sigma}(1)$ control the type I error in a conservative sense, they still achieve a high power.

\begin{table}[htb!]
\centering
\caption{The columns indexed with ``$CI_{\Sigma}(\tau)$" and ``$\phi_{\Sigma}(\tau)$" represent the empirical coverage of the constructed CIs and the empirical rejection rate of the testing procedure for $\beta_G^{\intercal}\Sigma_{G,G}\beta_G$ respectively. The columns indexed with ``$CI_{\mathrm I}(\tau)$" and ``$\phi_{\mathrm I}(\tau)$" represent the empirical coverage of the constructed CIs and the empirical rejection rate of the testing procedure for $\|\beta_G\|_2^2$ respectively.}
\label{tab: dense group} 
\scalebox{0.65}{
\begin{tabular}{@{}r|rrrrr|rrrrr@{}}
  \hline
 \multicolumn{11}{c}{Regression Setting ({\rm R1})}\\
 \hline
$n$ &$\beta_G^{\intercal}\Sigma_{G,G}\beta_G$ & $CI_{\Sigma}(\tau = 0)$ & $\phi_{\Sigma}(0)$ & $CI_{\Sigma}(\tau = 1)$ & $\phi_{\Sigma}(1)$ & $\beta_G^{\intercal}\beta_G$ & $CI_{\mathrm I}(\tau = 0)$ & $\phi_{\mathrm I}(0)$ & $CI_{\mathrm I}(\tau = 1)$ & $\phi_{\mathrm I}(1)$ \\
\hline
 350 & \multirow{3}{*}{0} 
 & 0.03 & 0.37 & 0.99 & 0.02 & \multirow{3}{*}{0} & 0.05 & 0.32 & 0.94 & 0.05 \\
 500 &  & 0.04 & 0.35 & 1.00 & 0.00 & & 0.08 & 0.34 & 0.97 & 0.03 \\
 800 &  & 0.08 & 0.34 & 1.00 & 0.00 & & 0.11 & 0.37 & 0.98 & 0.02 \\
\hline
\multicolumn{11}{c}{Regression Setting ({\rm R2})}\\
\hline
$n$ &$\beta_G^{\intercal}\Sigma_{G,G}\beta_G$ & $CI_{\Sigma}(\tau = 0)$ & $\phi_{\Sigma}(0)$ & $CI_{\Sigma}(\tau = 1)$ & $\phi_{\Sigma}(1)$ & $\beta_G^{\intercal}\beta_G$ & $CI_{\mathrm I}(\tau = 0)$ & $\phi_{\mathrm I}(0)$ & $CI_{\mathrm I}(\tau = 1)$ & $\phi_{\mathrm I}(1)$ \\
\hline
 350 & \multirow{3}{*}{0.023} 
 & 0.78 & 0.75 & 0.99 & 0.03 & \multirow{3}{*}{0.020} & 0.67 & 0.74 & 0.94 & 0.18 \\
 500 &  & 0.85 & 0.86 & 1.00 & 0.04 & & 0.69 & 0.76 & 0.95 & 0.26 \\
 800 &  & 0.89 & 0.90 & 1.00 & 0.05 & & 0.72 & 0.79 & 0.96 & 0.28 \\
 \hline
 \multicolumn{11}{c}{Regression Setting ({\rm R3})}\\
 \hline
 $n$ &$\beta_G^{\intercal}\Sigma_{G,G}\beta_G$ & $CI_{\Sigma}(\tau = 0)$ & $\phi_{\Sigma}(0)$ & $CI_{\Sigma}(\tau = 1)$ & $\phi_{\Sigma}(1)$ & $\beta_G^{\intercal}\beta_G$ & $CI_{\mathrm I}(\tau = 0)$ & $\phi_{\mathrm I}(0)$ & $CI_{\mathrm I}(\tau = 1)$ & $\phi_{\mathrm I}(1)$ \\
\hline
 350 & \multirow{3}{*}{0.866} 
 & 0.95 & 0.99 & 0.97 & 0.99 & \multirow{3}{*}{0.742} & 0.96 & 1.00 & 0.97 & 1.00 \\
 500 & & 0.93 & 1.00 & 0.96 & 1.00 &  & 0.96 & 1.00 & 0.97 & 1.00 \\
 800 & & 0.97 & 1.00 & 0.98 & 1.00 &  & 0.98 & 1.00 & 0.99 & 1.00 \\
 \hline
 \end{tabular}
}
\end{table}

\subsection{Mediation Analysis}
We study our inference procedure under a simulated high-dimensional mediation analysis setup. For $n \in {300, 600, 800}$, we generated exposure $T_i \sim N(0,2)$ for $i = 1, \ldots, n$ and $p \in \{500,1000\}$ potential mediators $M_i$ following $M_i = \gamma T_i + E_i$. Here, $\gamma$ is a $p$-dimensional vector such that $\gamma_1 = 0.20, \gamma_2 = 0.25, \gamma_3 = 0.15, \gamma_4 = 0.30, \gamma_5 = 0.35, \gamma_7 = 0.10$ and $\gamma_j = 0$ otherwise. $E_i \sim N(0,\Sigma_E)$ where $\Sigma_E = \{0.75^{|j - l|}\}_{1 \leq j,l \leq p}$. We have kept $\gamma$ fixed across replications. Finally, the count outcome is generated by $y_i \sim \text{Poisson}(e^{M_i^{\intercal}\beta_0 + T_i\beta_1})$. $\beta_0$ is generated as $\beta_{0,1} = 0.20, \beta_{0,2} = 0.25, \beta_{0,3} = 0.15, \beta_{0,4} = 0.30, \beta_{0,5} = 0.35, \beta_{0,6} = 0.10$ and $\beta_{0,j} = 0$ otherwise. Note that the first 5 of the $p$ potential mediators are the true mediators because both $\gamma$ and $\beta_0$ have non-zero entries corresponding to the first 5 indices. We fix the direct effect $\beta_1 = 0.5$. Finally we have $\gamma^{\intercal}\beta_0 = 0.3375$.  All results are averaged over 500 simulations. 
\vspace{2 mm}
\begin{table}[htb!]
\centering
\caption{The columns indexed by ``Cov", ``ERR", ``Bias" and ``SE" represent the empirical coverage of the CIs, empirical rejection rate of the proposed test, bias and standard error of the proposed indirect effect estimator respectively}
\label{tab: mediation-simu} 
\vspace{2 mm}
\scalebox{1.00}{
\begin{tabular}{@{}r|rrrr@{}}
  \hline
 \multicolumn{5}{c}{$p = 500$}\\
 \hline
$n$ & Cov & ERR & Bias & SE \\
\hline
 300 & 0.96 & 0.56 & 0.01 & 0.19 \\
 600 & 0.96 & 0.66 & -0.00 & 0.16 \\
 800 & 0.95 & 0.87 & -0.00 & 0.12 \\
 \hline
 \multicolumn{5}{c}{$p = 1000$}\\
 \hline
$n$ & Cov & ERR & Bias & SE \\
\hline
300 & 0.96 & 0.50 & 0.03 & 0.25 \\
600 & 0.95 & 0.61 & 0.01 & 0.18 \\
800 & 0.95 & 0.68 & 0.01 & 0.16 \\
\hline
 \end{tabular}
}
\end{table}

\vspace{2 mm}
\noindent Table~\ref{tab: mediation-simu} reports average coverage probabilities of our constructed CIs, the Empirical Rejection Rate (ERR) for our proposed testing procedure in \eqref{eq: testing med}, and the bias and standard error of the proposed indirect effect estimator in \eqref{eq: debias}. It can be observed that the constructed $95\%$ confidence intervals maintain the nominal coverage probability. The power of the test increases with the sample size. The standard error dominates the bias while being within acceptable limits. Overall, the performance appears to be acceptable.

\section{Real Data Application}
\label{sec: real data Poisson}
In this data set, we evaluate the effect of childhood maltreatment on post-traumatic stress disorder (PTSD) in adulthood, mediated by DNAms. It is hypothesized that childhood maltreatment affects biological processes via DNAm, which can have negative consequences later in life. However, in DNAm studies, the number of DNAm may far exceed the sample size, making high-dimensional mediation analysis necessary. Our method seeks to establish a causal connection between childhood maltreatment and post traumatic stress disorder (PTSD) and asses the mediating role of DNAm. We adopt the modified PTSD Symptom Scale (PSS) as the response variable; higher the PSS score, greater the probability of having comorbid PTSD and depression. The data used here is same as that analysed in \cite{PTSD}, originally from \cite{PTSDorg}. Logit transformed DNA methylation values from 335,669 
sites are used for analysis. We eliminate observations with missing values. 

\vspace{2 mm}
\noindent As pointed out in \cite{intermed1, intermed2} it is natural for the treatment to affect the response not only through the mediators but also through interacting with them. A mediation model including interaction between the exposure and the mediators can help to understand how childhood maltreatment can alter long-lasting DNAm changes which further affect psychological disorders such as PTSD. However, there exists high correlation among the genomic mediators. To address this issue, we select a subset of the mediators such that the maximum of the absolute correlation among them is below $0.7$. Eventually, we select a subset of $8955$ methylation loci. PSS score being count data, we fit the following Poisson outcome mediation model,

\begin{eqnarray}
    M_i & = & \gamma T_i + E_i; \quad i=1, \ldots, 403 \\ 
    y_i & \sim & \text{Poisson}\left(e^{M_i^{\intercal}\beta_0 +T_i\beta_1 + T_iM_i^{\intercal}\beta_2}\right),
    \label{eq: real med}
\end{eqnarray}

\noindent where for the $i-$th individual $M_i$ denotes the vector of mediators, $y_i$ refers to the PSS score while $T_i$ is an indicator whether the $i-$th individual suffered from childhood maltreatment.

\vspace{2 mm}
\noindent Defining $X_{i\cdot} := \left(M_i^{\intercal}, T_i, T_iM_i^{\intercal}\right)^{\intercal}$ and $\beta := \left(\beta_0, \beta_1, \beta_2\right)^{\intercal} \in \mathbb{R}^{17911}$, we can re-define the Poisson outcome model in \eqref{eq: real med} as $y_i \sim \text{Poisson}\left(e^{X_{i\cdot}^{\intercal}\beta}\right)$.

\vspace{2 mm}
\noindent Now the test of interaction $H_0 : \beta_2 = 0$ can be reformulated as $H_0 : \beta_\G = 0$ where $G = \{8861,\ldots,17911\}$. Using the hypothesis test in \eqref{eq: testing interaction} with $\tau = 0.5$, from Table \ref{tab: real data}, at 5$\%$ level of significance we found that childhood maltreatment and the DNA methylation do not interact to affect the response. However, applying the testing procedure in \eqref{eq: testing med} we found significant indirect effect of childhood maltreatment on PTSD mediated through the high-dimensional DNAm. This discovery opens up the possibility of finding mediators that are associated with the outcome. This information can be used to develop interventions that target these mediators. Note that the $M_j$s must be associated with both the output and the exposure to be a significant mediator. Methods developed in section \ref{sec: Inf linear} coupled with multiple testing corrections can be employed to answer such research questions. We leave it for future research.

\begin{table}[htb!]
\centering
\caption{``CI" stands for confidence interval for $\beta_G^{\intercal}\Sigma_{G,G}\beta_G$ in the first row and that for $\gamma^{\intercal}\beta_0$ in the second row.}
\label{tab: real data}
\scalebox{0.85}{
\begin{tabular}{@{}l|c|c@{}}
  \hline
Test & CI & Decision \\
  \hline
 Interaction  &  $[0,0.18]$ & No interaction between mediators and exposure. \\
\hline
 Indirect Effect & $[14.05,14.32]$  & There exists indirect effect of the exposure. \\
 \hline
 \end{tabular}
} 
\end{table}

\section{Discussion}
This paper has introduced a general framework for inference on both linear and quadratic functionals in high-dimensional Poisson regression. Our approach effectively handles the challenges of high-dimensional settings without imposing sparsity assumptions on the precision matrix or the loading vector, which sets it apart from previous work in the field.

\vspace{2 mm}
\noindent In our application to the epigenetic study, we examined PTSD score, a count variable, as the outcome. The results demonstrate the practical applicability of our methodology in analyzing real-world genomic data, particularly in the context of PTSD and its mediators.

\vspace{2 mm}
\noindent A key direction for future research is extending our framework to other count models, such as the negative binomial model, which would broaden its applicability to datasets where overdispersion is present and cannot be adequately captured by the Poisson assumption. Additionally, another open question is how to further refine our methodology to relax the boundedness assumption on the linear predictor $X_{i\cdot}^\intercal \beta$. Addressing this issue would enhance the flexibility of the framework and make it more robust in cases where such assumptions are restrictive.

\vspace{2 mm}
\noindent Finally, it is important to note that this paper is the first to systematically consider inference for both linear and quadratic functionals in the context of high-dimensional Poisson regression, laying the groundwork for future developments in this area. The extensive simulations and real-world application demonstrate the versatility and effectiveness of our proposed methods.

\bibliography{sample} 

\newpage

\section{Proofs}
We introduce the following events
{\small
$$
\mathcal{A}_1=\left\{\max_{1\leq i\leq n,\; 1\leq j\leq p}\left|X_{ij}\right|\leq C\sqrt{\log n+\log p}\right\}, \;\;
\mathcal{A}_2=\left\{\min_{\|\eta\|_2=1, \|\eta_{S^{c}}\|_1\leq C \|\eta_{S}\|_1}\frac{1}{n}\sum_{i=1}^{n}\left(X_{i\cdot}^{\intercal}\eta\right)^2\geq c\lambda_{\min}\left(\Sigma\right)\right\}$$
$$
\mathcal{A}_3=\left\{\min_{1\leq i\leq n}e^{X_{i\cdot}^{\intercal}\beta}\geq c_{\min}, \max_{1\leq i \leq n}e^{X_{i\cdot}^{\intercal}\beta} \leq c_{\max}\right\}, \;\;
\mathcal{A}_4=\left\{\left\|\frac{1}{n}\sum_{i=1}^{n}\epsilon_i X_{i\cdot}\right\|_{\infty}\leq C\sqrt{\frac{\log p}{n}}\right\}$$
$$\mathcal{A}_5=\left\{\|\widehat{\beta}-\beta\|_2\leq C \sqrt{\frac{k \log p}{n}}\right\},\;\;
\mathcal{A}_6=\left\{\|(\widehat{\beta}-\beta)_{S^{c}}\|_1\leq \|(\widehat{\beta}-\beta)_{S}\|_1\right\}$$}
\noindent where $S$ denotes the support of the high-dimensional vector $\beta$.
The following lemma \ref{lem: high prob a Poisson} controls the probability of these defined events and the proof is omitted as it is similar to Lemma 4 in \cite{cai2017confidence}. 
\begin{lemma} Suppose Conditions {\rm (C1)} holds, then 
$
\PP\left(\cap_{i=1}^{4}\mathcal{A}_i\right)\geq 1-g(n)-p^{-c}
$
and on the event $\cap_{i=1}^{4}\mathcal{A}_i$, the events $\mathcal{A}_5$ and $\mathcal{A}_6$ hold.
\label{lem: high prob a Poisson}
\end{lemma}

\subsection{Proof of Proposition \ref{prop: lasso convergence}}
For $t\in(0,1)$, by the definition of $\widehat{\beta}$, we have 
\begin{equation}
\ell(\widehat{\beta})+\lambda \|\widehat{\beta}\|_1\leq \ell(\widehat{\beta}+t(\beta-\widehat{\beta}))+\lambda \|\widehat{\beta}+t(\beta-\widehat{\beta})\|_1\leq \ell(\widehat{\beta}+t(\beta-\widehat{\beta}))+ (1-t) \lambda\|\widehat{\beta}\|_1+t\lambda\|\beta\|_1
\end{equation}
where $\ell(\beta)=\frac{1}{n} \sum_{i=1}^{n}\left(e^{X_{i\cdot}^{\intercal}\widehat{\beta}}-y_i\cdot\left(X_{i\cdot}^{\intercal}\beta\right)\right).$
Then we have 
\begin{equation}
\frac{\ell(\widehat{\beta})-\ell(\widehat{\beta}+t(\beta-\widehat{\beta}))}{t}+\lambda \|\widehat{\beta}\|_1\leq \lambda\|\beta\|_1 \quad \text{for any}\; t\in (0,1) 
\end{equation}
and taking the limit $t\to 0$, we have 
\begin{equation}
\frac{1}{n}\sum_{i=1}^{n}\left(e^{X_{i\cdot}^{\intercal}\widehat{\beta}}-y_i\right)X_{i\cdot}^{\intercal}(\widehat{\beta}-\beta)+\lambda \|\widehat{\beta}\|_1\leq \lambda\|\beta\|_1
\label{eq: basic inequality Poisson}
\end{equation}
Note that 
\begin{equation}
\begin{aligned}
&\left(e^{X_{i\cdot}^{\intercal}\widehat{\beta}}-y_i\right)X_{i\cdot}^{\intercal}(\widehat{\beta}-\beta)=\left(-\epsilon_i+\left(e^{X_{i\cdot}^{\intercal}\widehat{\beta}}-e^{X_{i\cdot}^{\intercal}\beta}\right)\right)X_{i\cdot}^{\intercal}(\widehat{\beta}-\beta)\\
&= - \epsilon_iX_{i\cdot}^{\intercal}(\widehat{\beta}-\beta)+ \int_{0}^{1} e^{X_{i\cdot}^{\intercal}\beta + tX_{i\cdot}^{\intercal}(\widehat{\beta}-\beta)}\left(X_{i\cdot}^{\intercal}(\widehat{\beta}-{\beta})\right)^2dt
\end{aligned}
\label{eq: lower bound 1 Poisson}
\end{equation}
Now we have 
\begin{equation}
\begin{aligned}
&e^{X_{i\cdot}^{\intercal}\beta + tX_{i\cdot}^{\intercal}(\widehat{\beta}-\beta)}\geq e^{X_{i\cdot}^{\intercal}\beta}\exp\left(-t\left|X_{i\cdot}^{\intercal}(\widehat{\beta}-{\beta})\right|\right)\\
&\geq e^{X_{i\cdot}^{\intercal}\beta}\exp\left(-t\max_{1\leq i\leq n}\left|X_{i\cdot}^{\intercal}(\widehat{\beta}-{\beta})\right|\right)
\end{aligned}
\label{eq: lower bound 2 Poisson}
\end{equation}
Combined with \eqref{eq: lower bound 1 Poisson}, we have 
\begin{equation}
\begin{aligned}
&\int_{0}^{1} e^{X_{i\cdot}^{\intercal}\beta + tX_{i\cdot}^{\intercal}(\widehat{\beta}-\beta)} \left(X_{i\cdot}^{\intercal}(\widehat{\beta}-{\beta})\right)^2dt\\
&\geq e^{X_{i\cdot}^{\intercal}\beta}\left(X_{i\cdot}^{\intercal}(\widehat{\beta}-{\beta})\right)^2 \int_{0}^{1}\exp\left(-t\max_{1\leq i\leq n}\left|X_{i\cdot}^{\intercal}(\widehat{\beta}-{\beta})\right|\right)d t\\
&=e^{X_{i\cdot}^{\intercal}\beta}\left(X_{i\cdot}^{\intercal}(\widehat{\beta}-{\beta})\right)^2 \frac{1-\exp\left(-\max_{1\leq i\leq n}\left|X_{i\cdot}^{\intercal}(\widehat{\beta}-{\beta})\right|\right)}{\max_{1\leq i\leq n}\left|X_{i\cdot}^{\intercal}(\widehat{\beta}-{\beta})\right|}
\end{aligned}
\end{equation}
Together with \eqref{eq: basic inequality Poisson}, we have 
\begin{equation}
\begin{aligned}
 &\frac{1-\exp\left(-\max_{1\leq i\leq n}\left|X_{i\cdot}^{\intercal}(\widehat{\beta}-{\beta})\right|\right)}{\max_{1\leq i\leq n}\left|X_{i\cdot}^{\intercal}(\widehat{\beta}-{\beta})\right|} \left(\frac{1}{n}\sum_{i=1}^{n}e^{X_{i\cdot}^{\intercal}\beta}\left(X_{i\cdot}^{\intercal}(\widehat{\beta}-{\beta})\right)^2\right)+\lambda \|\widehat{\beta}\|_1 \\
 &\leq \lambda \|\beta\|_1 + \frac{1}{n}\sum_{i=1}^{n}\epsilon_iX_{i\cdot}^{\intercal}(\widehat{\beta}-\beta)\leq \lambda \|\beta\|_1+\lambda_0\|\widehat{\beta}-\beta\|_1.
 \end{aligned}
 \end{equation}
 By the fact that $ \|\widehat{\beta}\|_1=\|\widehat{\beta}_{S}\|_1+\|\widehat{\beta}_{S^{c}}-\beta_{S^{c}}\|_1$ and $\|\beta\|_1-\|\widehat{\beta}_{S}\|_1\leq \|\widehat{\beta}_{S}-\beta_{S}\|_1$, then we have 
 \begin{equation}
 \begin{aligned}
& \frac{1-\exp\left(-\max_{1\leq i\leq n}\left|X_{i\cdot}^{\intercal}(\widehat{\beta}-{\beta})\right|\right)}{\max_{1\leq i\leq n}\left|X_{i\cdot}^{\intercal}(\widehat{\beta}-{\beta})\right|} \left(\frac{1}{n}\sum_{i=1}^{n}e^{X_{i\cdot}^{\intercal}\beta}\left(X_{i\cdot}^{\intercal}(\widehat{\beta}-{\beta})\right)^2\right)\\
&+\delta_0\lambda_0\|\widehat{\beta}_{S^{c}}-\beta_{S^{c}}\|_1\leq \left(2+\delta_0\right)\lambda_0\|\widehat{\beta}_{S}-\beta_{S}\|_1
 \end{aligned}
 \end{equation}
Then we deduce second inequality in \eqref{eq: est property} and
\begin{equation}
\begin{aligned}
& \frac{1-\exp\left(-\max_{1\leq i\leq n}\left|X_{i\cdot}^{\intercal}(\widehat{\beta}-{\beta})\right|\right)}{\max_{1\leq i\leq n}\left|X_{i\cdot}^{\intercal}(\widehat{\beta}-{\beta})\right|} \left(\frac{1}{n}\sum_{i=1}^{n}e^{X_{i\cdot}^{\intercal}\beta}\left(X_{i\cdot}^{\intercal}(\widehat{\beta}-{\beta})\right)^2\right)\\
 &\leq \left(2+\delta_0\right)\lambda_0\|\widehat{\beta}_{S}-\beta_{S}\|_1.
 \end{aligned}
 \label{eq: basic inequality a Poisson}
\end{equation}

\begin{lemma} On the event $\mathcal{A}_2\cap \mathcal{A}_3$, then 
\begin{equation}
\frac{1}{n}\sum_{i=1}^{n}e^{X_{i\cdot}^{\intercal}\beta}\left(X_{i\cdot}^{\intercal}(\widehat{\beta}-{\beta})\right)^2\geq c_{\min}\lambda_{\min}\left(\Sigma\right)\|\widehat{\beta}-{\beta}\|_2^2
\end{equation}
\end{lemma}
\noindent Then \eqref{eq: basic inequality a Poisson} is further simplified as 
\begin{equation}
 \frac{1-\exp\left(-\max_{1\leq i\leq n}\left|X_{i\cdot}^{\intercal}(\widehat{\beta}-{\beta})\right|\right)}{\max_{1\leq i\leq n}\left|X_{i\cdot}^{\intercal}(\widehat{\beta}-{\beta})\right|} c_{\min}\lambda_{\min}\left(\Sigma\right)\|\widehat{\beta}-{\beta}\|_2^2\leq  \left(2+\delta_0\right)\lambda_0\|\widehat{\beta}_{S}-\beta_{S}\|_1
  \label{eq: basic inequality c Poisson}
\end{equation}
{\bf Case 1:} Assume that 
\begin{equation}
\max_{1\leq i\leq n}\left|X_{i\cdot}^{\intercal}(\widehat{\beta}-{\beta})\right|\leq c_1 \quad \text{for some}\; c_1>0 
\label{eq: assisting assumption Poisson}
\end{equation} then we have 
\begin{equation}
\begin{aligned}
&\frac{1-\exp\left(-\max_{1\leq i\leq n}\left|X_{i\cdot}^{\intercal}(\widehat{\beta}-{\beta})\right|\right)}{\max_{1\leq i\leq n}\left|X_{i\cdot}^{\intercal}(\widehat{\beta}-{\beta})\right|}=\int_{0}^{1}\exp\left(-t\max_{1\leq i\leq n}\left|X_{i\cdot}^{\intercal}(\widehat{\beta}-{\beta})\right|\right)d t\\
&\geq \int_{0}^{1}\exp\left(-t c_1\right)dt=\frac{1-\exp\left(-c_1\right)}{c_1}
\end{aligned}
\end{equation}
Define $c_2=\frac{c_{\min}\lambda_{\min}\left(\Sigma\right)}{2+\delta_0} \frac{1-\exp\left(-c_1\right)}{c_1}$, then we have
\begin{equation}
c_2 \|\widehat{\beta}-{\beta}\|_2^2\leq \lambda_0\|\widehat{\beta}_{S}-\beta_{S}\|_1\leq \sqrt{k}\lambda_0 \|\widehat{\beta}_{S}-\beta_{S}\|_2
\end{equation}
and hence 
\begin{equation}
\|\widehat{\beta}-\beta\|_2\lesssim \frac{1}{\lambda_{\min}}\sqrt{k}\lambda_0 \quad \text{and} \quad \|\widehat{\beta}-\beta\|_1\leq {k}\lambda_0
\label{eq: error bound under assumption Poisson}
\end{equation}
{\bf Case 2:}  Assume that \eqref{eq: assisting assumption Poisson} does not hold, then 
\begin{equation}
\frac{1-\exp\left(-\max_{1\leq i\leq n}\left|X_{i\cdot}^{\intercal}(\widehat{\beta}-{\beta})\right|\right)}{\max_{1\leq i\leq n}\left|X_{i\cdot}^{\intercal}(\widehat{\beta}-{\beta})\right|}\geq \frac{1-\exp(-c_1)}{\max_{1\leq i\leq n}\left|X_{i\cdot}^{\intercal}(\widehat{\beta}-{\beta})\right|}
\end{equation}
Together with \eqref{eq: basic inequality c Poisson}, we have
\begin{equation}
c_2 c_1\|\widehat{\beta}-{\beta}\|_2^2\leq \lambda_0\|\widehat{\beta}_{S}-\beta_{S}\|_1\max_{1\leq i\leq n}\left|X_{i\cdot}^{\intercal}(\widehat{\beta}-{\beta})\right|
\label{eq: bound 1 Poisson}
\end{equation}
By $\max_{1\leq i\leq n}\left|X_{i\cdot}^{\intercal}(\widehat{\beta}-{\beta})\right|\leq  \max\left|X_{ij}\right|\|\widehat{\beta}-\beta\|_1$ and \eqref{eq: est property}, we further have
\begin{equation}
\begin{aligned}
\lambda_0\|\widehat{\beta}_{S}-\beta_{S}\|_1\max_{1\leq i\leq n}\left|X_{i\cdot}^{\intercal}(\widehat{\beta}-{\beta})\right|&\leq \frac{2+2\delta_0}{\delta_0}  \max\left|X_{ij}\right| \lambda_0\|\widehat{\beta}_{S}-\beta_{S}\|_1^2\\
&\leq  \frac{2+2\delta_0}{\delta_0} \max\left|X_{ij}\right| k \lambda_0\|\widehat{\beta}_{S}-\beta_{S}\|_2^2,
\end{aligned}
\label{eq: bound 2 Poisson}
\end{equation}
where the last inequality follows from Cauchy inequality. Combining \eqref{eq: bound 1 Poisson} and \eqref{eq: bound 2 Poisson}, we have shown that if \eqref{eq: assisting assumption Poisson} does not hold, then \begin{equation}
\max\left|X_{ij}\right| \frac{2+2\delta_0}{\delta_0} k \lambda_0\geq c_2 c_1,
\end{equation}
Since this contradicts the assumption that $\max\left|X_{ij}\right| k\lambda_0< \frac{c_2 c_1\delta_0}{2+2\delta_0}$, we establish \eqref{eq: error bound under assumption Poisson} and hence  \eqref{eq: est property}.

\section{Proof of Lemma \ref{lem: variance lower bound}} 
On the event $\mathcal{A}_3$, we have
$
\frac{1}{c_{\max}}\widehat{u}^{\intercal}\left[\frac{1}{n^{2}}\sum_{i=1}^{n}X_{i\cdot}X_{i\cdot}^{\intercal}\right]\widehat{u} \leq {\rm V}_{x_*} \leq \frac{1}{c_{\min}} \widehat{u}^{\intercal}\left[\frac{1}{n^{2}}\sum_{i=1}^{n}X_{i\cdot}X_{i\cdot}^{\intercal}\right]\widehat{u}.$ The rest of the proof follows exactly as in the proof of (20) in \cite{caseprob}.
\subsection{Proof of Theorem \ref{thm: limiting distribution}}
Now, recall the error decomposition of the proposed estimator $\widehat{x_*^{\intercal}\beta}$ in \eqref{eq: error decomposition}.  The following proposition controls the second and third terms.
\begin{proposition} Suppose that Condition \textrm{(C1)} holds. For any estimator $\widehat{\beta}$ satisfying Condition \eqref{eq: est property}, with probability larger than $1-p^{-c}-g(n)$ for some positive constant $c>0$,
	\begin{equation}
	n^{1/2}\left|(\widehat{\Sigma}\widehat{u}-\xnew)^{\intercal}(\widehat{\beta}-\beta)\right|\leq n^{1/2}\|\xnew\|_2\lambda_{n}\|\widehat{\beta}-\beta\|_1\lesssim \|\xnew\|_2 {k \log p}\cdot{n^{-1/2}},
	\label{eq: error bound 1 Poisson}
	\end{equation}
	and 
	\begin{equation}
	n^{1/2}|\widehat{u}^{\intercal}\frac{1}{n}\sum_{i=1}^{n}e^{-X_{i\cdot}^{\intercal}\widehat{\beta}}X_{i\cdot}\Delta_i|\leq  \eta_{n} \|\xnew\|_2 {k \log p}\cdot{n^{-1/2}}
	\label{eq: error bound 3 Poisson}
	\end{equation}
	\label{prop: decomposition Poisson}
\end{proposition}
 \underline{\bf Proof} : 

 \noindent \underline{Proof of \eqref{eq: error bound 1 Poisson}} The first inequality of \eqref{eq: error bound 1 Poisson} follows from Holder's inequality and the second inequality follows from Condition {\eqref{eq: est property}}.
 
\noindent \underline{Proof of \eqref{eq: error bound 3 Poisson}} Note that by assumption \textrm{(C1)} on the event $\mathcal{A}=\cap_{i=1}^{6}\mathcal{A}_{i}$,
$$\left|X_{i\cdot}^{\intercal}\widehat{\beta}\right| \leq \left|X_{i\cdot}^{\intercal}\beta\right| + \left|X_{i\cdot}^{\intercal}(\widehat{\beta}-\beta)\right| \leq C + C\sqrt{\log n + \log p}.k\sqrt{\frac{\log p}{n}} \leq C \quad \forall i$$ for some constant $C > 0$ and the last inequality follows from the fact that $\sqrt{n}\gg k\log p\left(1+\sqrt{\frac{\log n}{\log p}}\right).$. Therefore, $
\sum_{i=1}^{n}\left|\Delta_i\right|\leq C.\sum_{i=1}^{n}\left(X^{\intercal}_{i\cdot}(\widehat{\beta}-\beta)\right)^2$ and $e^{-X_{i\cdot}^{\intercal}\widehat{\beta}} \leq e^C$.
\noindent By Cauchy inequality, we have 
\begin{equation}
\sqrt{n}\left|\widehat{u}^{\intercal}\frac{1}{n}\sum_{i=1}^{n}e^{-X_{i\cdot}^{\intercal}\widehat{\beta}}X_{i\cdot}\Delta_i\right|\leq \max_{1\leq i\leq n}\left|\widehat{u}^{\intercal}X_{i\cdot}\right|\frac{1}{\sqrt{n}}\sum_{i=1}^{n} e^{-X_{i\cdot}^{\intercal}\widehat{\beta}}\left|\Delta_i\right|\lesssim C_1\eta_{n} \|\xnew\|_2 \sum_{i=1}^{n}\left(X_{i\cdot}^{\intercal}(\widehat{\beta}-\beta)\right)^2
\label{eq: diminishing error b Poisson}
\end{equation}
for $C_1 := C.e^C$. On the event $\mathcal{A}$, we have 
\begin{equation}
\frac{1}{n}\sum_{i=1}^{n}\left(X^{\intercal}_{i\cdot}(\widehat{\beta}-\beta)\right)^2\leq C \|\widehat{\beta}-\beta\|_2^2\leq C\frac{k \log p}{n}.
\label{eq: inter bound 1 Poisson}
\end{equation}
Together with \eqref{eq: diminishing error b Poisson} and \eqref{eq: inter bound 1 Poisson}, we establish that, on the event $\mathcal{A}$,
\begin{equation}
\sqrt{n}\left|\widehat{u}^{\intercal}\frac{1}{n}\sum_{i=1}^{n}e^{-X_{i\cdot}^{\intercal}\widehat{\beta}}X_{i\cdot}\Delta_i\right| \leq C \eta_{n} \|\xnew\|_2 \frac{k \log p}{\sqrt{n}}.
\end{equation}

\noindent Together with \eqref{eq: error decomposition}, it remains to establish the asymptotic normality of the following term,
\begin{equation}
\widehat{u}^{\intercal}\frac{1}{n}\sum_{i=1}^{n} e^{-X_{i\cdot}^{\intercal}\widehat{\beta}}X_{i\cdot}\epsilon_i.
\label{eq: reweight-term Poisson}
\end{equation}
The technical complexity in establishing the asymptotic normality of this reweighted summation \eqref{eq: reweight-term Poisson} is the dependence between the weight $e^{-X_{i\cdot}^{\intercal}\widehat{\beta}}$ and the model error $\epsilon_i$. The correlation between $\widehat{\beta}$ and $\epsilon_i$ is decoupled through the following decomposition,
\begin{equation}
\widehat{u}^{\intercal}\frac{1}{n}\sum_{i=1}^{n} e^{-X_{i\cdot}^{\intercal}\widehat{\beta}}X_{i\cdot}\epsilon_i=\widehat{u}^{\intercal}\frac{1}{n}\sum_{i=1}^{n} e^{-X_{i\cdot}^{\intercal}\beta}X_{i\cdot}\epsilon_i\\
+{\widehat{u}}^{\intercal}\frac{1}{{n}}\sum_{i=1}^{n}\left(e^{-X_{i\cdot}^{\intercal}\widehat{\beta}}-e^{-X_{i\cdot}^{\intercal}\beta}\right)X_{i\cdot}\epsilon_i.
\label{eq: technical decomposition Poisson}
\end{equation}
The second component on the right hand side of \eqref{eq: technical decomposition Poisson} captures the error incurred on estimating $\beta$ by $\widehat{\beta}$. We now provide a sharp control of this error term by suitable empirical process theory in Proposition \ref{prop: bias control} and the asymptotic normality of the first term in \eqref{eq: technical decomposition Poisson} is proved in Proposition \ref{prop: dominating variance}. 

\begin{proposition}
Suppose that Condition \textrm{(C1)} holds and the initial estimator $\widehat{\beta}$ satisfies Condition \eqref{eq: est property}, then with probability greater than $1-p^{-c}-g(n)-{1}/{t_0}$,
\begin{equation}
\left|{\widehat{u}}^{\intercal}\frac{1}{{n}}\sum_{i=1}^{n}\left(e^{-X_{i\cdot}^{\intercal}\widehat{\beta}}-e^{-X_{i\cdot}^{\intercal}\beta}\right)X_{i\cdot}\epsilon_i\right| \leq C t_0 \eta_n \|\xnew\|_2\frac{k \log p \log n}{n} 
\label{eq: error bound 2 Poisson}
\end{equation}
where $\eta_{n}$ is defined in \eqref{eq: constraint 3 Poisson}, $t_0>1$ is a large positive constant and $c>0$ and $C>0$ are positive constants.
\label{prop: bias control}
\end{proposition}
\underline{\bf Proof} : To facilitate the proof, we introduce the following two lemmas

\begin{lemma}
Suppose that $y_i'$ is an independent copy of $y_i$ and $\epsilon'_i$ is defined as $y'_i-\E(y'_i\mid X_{i\cdot})$. For all convex non-decreasing functions $\Phi: \R_{+}\to \R_{+}$, then 
\begin{equation}
\E \Phi\left(\sup_{t\in \mathcal{T}}\left|\sum_{i=1}^{n} g_i(t_i) \epsilon_i\right|\right)\leq \E \Phi\left(\sup_{t\in \mathcal{T}}\left|\sum_{i=1}^{n} g_i(t_i) \xi_i\right|\right),
\label{eq: symmetrization Poisson}
\end{equation}
where 
$
\xi_i=\epsilon_i-\epsilon^{\prime}_i=y_i-y^{\prime}_i.
$
\label{lem: symmetrization Poisson}
\end{lemma}
 \underline{\bf Proof} : 
    The proof is same as that of Lemma 6 in \cite{caseprob}.

\begin{lemma} Let $t=(t_1,\cdots,t_n)\in\mathcal{T}\subset R^{n}$ and let $\phi_i: \R\to\R, i=1,\cdots,n$ be functions such that $\phi_i(0)=0$ and 
$
\left|\phi_i(u)-\phi_i(v)\right|\leq |u-v|, u,v\in \R.
$
For all convex non-decreasing functions $\Phi: \R_{+}\to \R_{+}$, then 
\begin{equation}
\E \Phi\left(\frac{1}{2}\sup_{t\in \mathcal{T}}\left|\sum_{i=1}^{n}\phi_i(t_i)\xi_i\right|\right)\leq \E \Phi\left(\sup_{t\in \mathcal{T}}\left|\sum_{i=1}^{n}t_i\xi_i\right|\right),
\label{eq: contraction Poisson}
\end{equation}
where $\left\{\xi_i\right\}_{1\leq i\leq n}$ are independent random variables with the probability density function 
\begin{equation}
\PP\left(\xi_i=k\right)=\PP\left(\xi_i=-k\right)\in [0, \frac{1}{2}]\; \forall k = 1,2,\cdots\text{and}\; \; \PP\left(\xi_i=0\right)=1-2\sum_{k=1}^{\infty}\PP\left(\xi_i=k\right).
\label{eq: distribution Poisson}
\end{equation}
\label{lem: contraction Poisson}
\end{lemma}
 \underline{\bf Proof} : 
This lemma is proved for binary outcome GLMs in \citep{caseprob}. We need to extend it to the case of Poisson regression. In case of Poisson regression, with $\pi = X_{i\cdot}^{\intercal}\beta, \PP(\epsilon_i = k - e^{\pi}) = \frac{e^{\pi}\pi^{k}}{k!}$. Therefore $\xi_i = \epsilon_i - \epsilon^{\prime}_i$ takes values in $ \{\ldots, -2, -1, 0, 1, 2, \ldots\}$ such that \eqref{eq: distribution Poisson} is satisfied. 

\noindent The proof follows from that of Theorem 2.2 in \cite{koltchinskii2011oracle} and some modification is necessary to extend the results to the general random variables $\xi_1, \xi_2,\cdots, \xi_{n}$ which are independent and follow the probability distribution \eqref{eq: distribution Poisson}.

\noindent We start with proving the following inequality for a function $A: \mathcal{T}\to \R,$
\begin{equation}
\E \Phi\left(\sup_{t\in \mathcal{T}}[A(t)+\sum_{i=1}^{n}\phi_i(t_i)\xi_i]\right)\leq \E\Phi\left(\sup_{t\in \mathcal{T}}[A(t)+\sum_{i=1}^{n} t_i\xi_i]\right),
\label{eq: inter contraction Poisson}
\end{equation}
We first prove the special case $n=1$, which is reduced to be the following inequality,
\begin{equation}
\E \Phi\left(\sup_{t\in \mathcal{T}}[t_1+\phi(t_2)\xi_0]\right)\leq \E \Phi\left(\sup_{t\in \mathcal{T}}[t_1+t_2\xi_0]\right),
\label{eq: n=1 Poisson}
\end{equation}
where $\mathcal{T}\subset \R^2$. In that spirit we prove that $\forall k \in \left\{\ldots, -1,0,1, \ldots\right\}$
\begin{equation*}
    \Phi\left(\sup_{t \in \mathcal{T}}[t_1 + k\phi(t_2)]\right) + \Phi\left(\sup_{t \in \mathcal{T}}[t_1 - k\phi(t_2)]\right) \leq \Phi(\sup_{t\in \mathcal{T}}[t_1 + kt_2]) + \Phi(\sup_{t\in \mathcal{T}}[t_1 - kt_2])
\end{equation*}
The above inequality follows from the same line of proof as that in \cite{koltchinskii2011oracle}. Then
\small{\begin{equation}
    \begin{aligned}
        \mathbb{E}\Phi\left(\sup_{t \in \mathcal{T}}[t_1 + \phi(t_2)\xi_0]\right) & = \sum_{k = -\infty}^{\infty}\mathbb{P}(\xi_0 = k)\Phi\left(\sup_{t \in \mathcal{T}}[t_1 + \phi(t_2)k]\right) \\
        & = \sum_{k = 1}^{\infty} \mathbb{P}(\xi_0 = k)\left[\Phi\left(\sup_{t \in \mathcal{T}}[t_1 + \phi(t_2)k]\right) + \Phi\left(\sup_{t \in \mathcal{T}}[t_1 - \phi(t_2)k]\right) \right]+\mathbb{P}(\xi_0 = 0)\Phi\left(\sup_{t \in \mathcal{T}} t_1\right) \\
        & \leq \sum_{k = 0}^{\infty} \mathbb{P}(\xi_0 = k)\left[\Phi\left(\sup_{t \in \mathcal{T}}[t_1 + t_2k]\right) + \Phi\left(\sup_{t \in \mathcal{T}}[t_1 - t_2k]\right)\right]+\mathbb{P}(\xi_0 = 0)\Phi\left(\sup_{t \in \mathcal{T}} t_1\right) \\
        & = \sum_{k = \infty}^{\infty}\mathbb{P}(\xi_0 = k)\Phi\left(\sup_{t \in \mathcal{T}}[t_1 + t_2k]\right) \\
        & = \mathbb{E}\Phi\left(\sup_{t \in \mathcal{T}}[t_1 + t_2\xi_0]\right)
    \end{aligned}
\end{equation}}
The second equality follows since $\mathbb{P}(\xi_0 = k) = \mathbb{P}(\xi_0 = -k)$. It remains to prove the lemma by applying induction and \eqref{eq: n=1 Poisson}, that is,
\begin{equation*}
\begin{aligned}
&\E_{ ({\xi}_1,\cdots, \xi_{n}) \mid X} \Phi\left(\sup_{t\in \mathcal{T}}[A(t)+\sum_{i=1}^{n}\phi_i(t_i)\xi_i]\right)= \E_{ ({\xi}_1,\cdots, \xi_{n-1}) \mid X} \E_{ \xi_{n} \mid X} \Phi\left(\sup_{t\in \mathcal{T}}[A(t)+\sum_{i=1}^{n}\phi_i(t_i)\xi_i]\right)\\
&\leq \E_{ ({\xi}_1,\cdots, \xi_{n-1}) \mid X} \E_{ \xi_{n} \mid X} \Phi\left(\sup_{t\in \mathcal{T}}[A(t)+\sum_{i=1}^{n-1}\phi_i(t_i)\xi_i+t_{n}\xi_{n}]\right)\\
&=\E_{ \xi_{n} \mid X}\E_{ ({\xi}_1,\cdots, \xi_{n-1}) \mid X}  \Phi\left(\sup_{t\in \mathcal{T}}[A(t)+\sum_{i=1}^{n-1}\phi_i(t_i)\xi_i+t_{n}\xi_{n}]\right)
\end{aligned}
\end{equation*}
Continuing the above equation, we establish $$\E_{ ({\xi}_1,\cdots, \xi_{n}) \mid X} \Phi\left(\sup_{t\in \mathcal{T}}[A(t)+\sum_{i=1}^{n}\phi_i(t_i)\xi_i]\right)\leq \E_{ ({\xi}_1,\cdots, \xi_{n}) \mid X} \Phi\left(\sup_{t\in \mathcal{T}}[A(t)+\sum_{i=1}^{n} t_i\xi_i]\right)$$. Integration with respect to $X$ leads to \eqref{eq: inter contraction Poisson}. In the following, we will apply \eqref{eq: inter contraction Poisson} to establish \eqref{eq: contraction Poisson}. Note that 
\begin{equation*}
\begin{aligned}
&\E \Phi\left(\frac{1}{2}\sup_{t\in \mathcal{T}}\left|\sum_{i=1}^{n}\phi_i(t_i)\xi_i\right|\right)=\E \Phi\left(\frac{1}{2}\left(\sup_{t\in \mathcal{T}}\sum_{i=1}^{n}\phi_i(t_i)\xi_i\right)_{+}+\frac{1}{2}\left(\sup_{t\in \mathcal{T}}\sum_{i=1}^{n}\phi_i(t_i)(-\xi_i)\right)_{+}\right)\\
&\leq\frac{1}{2}\left[\E \Phi\left(\left(\sup_{t\in \mathcal{T}}\sum_{i=1}^{n}\phi_i(t_i)\xi_i\right)_{+}\right)+\E \Phi\left(\left(\sup_{t\in \mathcal{T}}\sum_{i=1}^{n}\phi_i(t_i)(-\xi_i)\right)_{+}\right)\right]
\end{aligned}
\end{equation*}
By applying \eqref{eq: inter contraction Poisson} to the function $u\to \Phi(u_{+})$, which is convex and non-decreasing, we have 

\noindent $\E \Phi\left(\left(\sup_{t\in \mathcal{T}}\sum_{i=1}^{n}\phi_i(t_i)\xi_i\right)_{+}\right)\leq \E\Phi\left(\sup_{t\in \mathcal{T}}\sum_{i=1}^{n}t_i\xi_i\right)\leq \E\Phi\left(\sup_{t\in \mathcal{T}}\left|\sum_{i=1}^{n} t_i \xi_i\right|\right)$. Then we establish \eqref{eq: contraction Poisson}.
 
\noindent We now come back to proof of Proposition \ref{prop: bias control}. Define $g_i(t_i) := \left(\frac{1}{e^{X_{i\cdot}^{\intercal}\beta + t_i}} - \frac{1}{e^{X_{i\cdot}^{\intercal}\beta}}\right)\widehat{u}^{\intercal}X_{i\cdot}$ and the space for $\delta\in \R^{p}$ as
 \begin{equation}
 \mathcal{C}=\left\{\delta: \|\delta_{S^{c}}\|_1\leq  \|\delta_{S}\|_1,\; \|\delta\|_2\leq C^{*} \sqrt{\frac{k \log p}{n}}\right\}.
 \label{eq: restricted space Poisson}
 \end{equation}
 for some positive constant $C^{*}>0$.
We further define 
\begin{equation}
\mathcal{T}=\left\{t=(t_1,\cdots,t_n): t_i=X_{i\cdot}^{\intercal}\delta \;\;\text{where}\;\; \delta \in \mathcal{C}\right\},
\label{eq: mapped restricted space Poisson}
\end{equation}
Then we can re-write
\begin{eqnarray}
     \sqrt{n} \left|\widehat{u}^{\intercal}\frac{1}{n}\sum_{i=1}^{n}\left(\frac{1}{e^{X_{i\cdot}^{\intercal}\widehat{\beta}}} - \frac{1}{e^{X_{i\cdot}^{\intercal}\beta}}\right)X_{i\cdot}\epsilon_i\right| & \leq & \sup_{\delta\in \mathcal{C}}\left|\frac{1}{\sqrt{n}}\sum_{i=1}^{n} g_i(X_{i\cdot}^{\intercal}\delta)\cdot \epsilon_i\right| \\
     & = & \sup_{t\in \mathcal{T}}\left|\frac{1}{\sqrt{n}}\sum_{i=1}^{{n}} g_i(t_i)\cdot \epsilon_i\right| \label{eq: bound over sup a Poisson}
\end{eqnarray}
 For $t,s\in \mathcal{T} \subset \R^{n}$, then there exist $\delta^{t},\delta^{s}\in \mathcal{C}\subset \R^{p}$ such that 
$
t_i-s_i=X_{i\cdot}^{\intercal}\left(\delta^{t}-\delta^{s}\right)$ and $t_i=X_{i\cdot}^{\intercal}\delta^{t}$ for $1\leq i\leq n.$
Hence on the event $\mathcal{A}_1,$
\begin{equation}
\max\left\{\max_{1\leq i\leq n}|t_i-s_i|,\max_{1\leq i\leq n}|t_i|\right\}\leq C k \sqrt{\frac{\log p}{n}} \sqrt{\log p+\log n}\leq 1. 
\label{eq: difference control Poisson}
\end{equation}
where the last inequality follows as long as $\sqrt{n} \geq k \log p \left(1+\sqrt{\frac{\log n}{\log p}}\right)$ 

\noindent Define $q(x) := e^{-x}$ and then 
\begin{eqnarray}
g_i(s_i)-g_{i}(t_i)
&=&\left(q\left(X_{i\cdot}^{\intercal}{\beta}+s_i\right) - q\left(X_{i\cdot}^{\intercal}{\beta}+t_i\right)\right)\widehat{u}^{\intercal}X_{i\cdot}.\\
&=& q^{\prime}(X_{i\cdot}^{\intercal}\beta + \widetilde{t}_i)(s_i - t_i)\widehat{u}^{\intercal}X_{i\cdot} \\
& = & e^{-(X_{i\cdot}^{\intercal}\beta + \widetilde{t}_i)}(s_i - t_i)\widehat{u}^{\intercal}X_{i\cdot}
\label{eq: re-express Poisson}
\end{eqnarray}
where $\widetilde{t}_i$ lies between $s_i$ and $t_i$. Observe, since $t_i \in \mathcal{C}$, on $\mathcal{A}$
\begin{eqnarray*}
    |t_i| & \leq & \max_{1\leq i \leq n, 1 \leq j \leq p}|X_{i,j}|\|\delta\|_1 \\
    & \leq & \sqrt{\log p + \log n}.k\sqrt{\frac{\log p}{n}} \\
    & \leq & C
\end{eqnarray*}
for some constant $C>0$ because we assume $\sqrt{n} \geq k \log p (1 + \sqrt{\frac{\log n}{\log p}})$. This implies $|\widetilde{t}_i|\leq C$. Therefore by assumption \textrm{(C1)}, $e^{-(X_{i\cdot}^{\intercal}\beta + \widetilde{t}_i)} \leq C_1$ for some $C_1 > 0$. Hence \eqref{eq: re-express Poisson} says that 
$$
|g_{i}(s_i) - g_i(t_i)| \leq C_1\eta_n\|x_*\|_2|s_i - t_i|
$$
Let $\phi_i(u) = \frac{g_i(u)}{L_n}$ with $L_n = C_1\eta_n\|x_*\|_2$ and $\Phi(x) = x$. Now we apply Lemmas \ref{lem: symmetrization Poisson} and \ref{lem: contraction Poisson} to get
\begin{equation*}
\E_{\xi\mid X} \sup_{t\in \mathcal{T}}\left|\frac{1}{n}\sum_{i=1}^{n} \phi_i(t_i)\cdot \xi_i\right|\leq 2\E_{\xi\mid X} \sup_{\delta\in \mathcal{C}} \left|\frac{1}{n}\sum_{i=1}^{n}\delta^{\intercal}X_{i\cdot}\xi_i\right|
\label{eq: contraction app1 Poisson}
\end{equation*}
and hence 
$
\E\sup_{t\in \mathcal{T}}\left|\frac{1}{n}\sum_{i=1}^{n} \phi_i(t_i)\cdot\xi_i\right|\leq 2\E \sup_{\delta\in \mathcal{C}} \left|\frac{1}{n}\sum_{i=1}^{n}\delta^{\intercal}X_{i\cdot}\xi_i\right|.$
Note that
\begin{equation*}
\E \sup_{\delta\in \mathcal{C}} \left|\frac{1}{n}\sum_{i=1}^{n}\delta^{\intercal}X_{i\cdot}\xi_i\right|\leq \sup_{\delta\in \mathcal{C}}\|\delta\|_1 \E \left\|\frac{1}{n}\sum_{i=1}^{n}X_{i\cdot}\xi_i\right\|_{\infty}
\end{equation*}
Now, recall $\mathcal{A}_1 = \left\{\max_{i,j}|X_{i,j}| \leq C_1\sqrt{\log p + \log n}\right\}$. Define $\widetilde{\mathcal{A}} := \left\{\max_{i}|\xi_{i}| \leq C_2\sqrt{\log n}\right\}$ where $C_1, C_2 > 0$ are some constants. Since conditional on $\{X_{i\cdot}\}_{i=1}^{n}$, $\xi_i$ is a sub-exponential random variable, $\mathbb{P}(\widetilde{\mathcal{A}}) \geq 1 - n^{-c}$ for some constant $c>0$. Note that, on $\mathcal{A}_1\cap\widetilde{\mathcal{A}}$, we have $\max_{i,j}\left|X_{ij}\xi_i\right| \leq C\sqrt{\log n}\sqrt{\log n + \log p}$. By applying Hoeffding's inequality we have
\begin{equation*}
    \mathbb{P}\left(\left|\frac{1}{n}\sum_{i=1}^{n}X_{ij}\xi_i\right| \geq \frac{(\log n)\sqrt{\log p}}{\sqrt{n}}\right) \lesssim n^{-c}
\end{equation*}
Therefore $$\E \left\|\frac{1}{n}\sum_{i=1}^{n}X_{i\cdot}\xi_i\right\|_{\infty} = \E\max_{1\leq j \leq p}\left|\frac{1}{n}\sum_{i=1}^{n}X_{i,j}\xi_i\right| \lesssim \log n \sqrt{\frac{\log p}{n}}$$.
Combined with  $\sup_{\delta\in \mathcal{C}}\|\delta\|_1\leq C^* k\sqrt{\frac{\log p}{n}}$, we establish 

\noindent $\E \sup_{\delta\in \mathcal{C}} \left|\frac{1}{n}\sum_{i=1}^{n}\delta^{\intercal}X_{i\cdot}\xi_i\right|\leq C \frac{k \log p \log n}{n}$ and $
\E\sup_{t\in \mathcal{T}}\left|\frac{1}{n}\sum_{i=1}^{n} \phi_i(t_i)\cdot {\bf 1}_{\mathcal{A}_1} \xi_i\right|\leq C \frac{k \log p \log n}{n}.
$
By Chebyshev's inequality,  
$$
\PP\left(\sup_{t\in \mathcal{T}}\left|\frac{1}{n}\sum_{i=1}^{n} g_i(t_i)\cdot {\bf 1}_{\mathcal{A}_1} \xi_i\right|\geq  C t \|\xnew\|_2 \eta_n \frac{k \log p \log n}{n}
\right)\leq \frac{1}{t}.
$$
By \eqref{eq: bound over sup a Poisson}, we establish that  \eqref{eq: error bound 2 Poisson} holds with probability larger than $1-(\frac{1}{t}+p^{-c}+g(n)).$

\begin{proposition} Suppose that Condition \textrm{(C1)} holds and $\eta_{n}$ defined in \eqref{eq: constraint 3 Poisson} satisfies $\left(\log n\right)^{1/2}\lesssim \eta_{n}\ll n^{1/2}$, then
\begin{equation}
    \frac{1}{\mathrm{V}_{\xnew}^{1/2}}\widehat{u}^{\intercal}\frac{1}{n}\sum_{i=1}^{n} e^{-X_{i\cdot}^{\intercal}\beta}X_{i\cdot}\epsilon_i\to N(0,1)
    \label{eq: central limit Poisson}
\end{equation}
where ${\rm V}_{\xnew}$ is defined in \eqref{eq: asymptotic variance}.
\label{prop: dominating variance}
\end{proposition}
\underline{\bf Proof} : 
Let $W_i := \frac{1}{n} \frac{\widehat{u}^{\intercal}e^{-X_{i\cdot}^{\intercal}\beta}X_{i\cdot}\epsilon_i}{\sqrt{\mathrm{V}_{\xnew}}}$. Conditioning on $X$, then $\{W_i\}_{1\leq i\leq n}$ are independent random variables with $\E(W_i\mid X_{i\cdot})=0$ and $\sum_{i=1}^{n}{\rm Var}(W_i\mid X_{i\cdot})=1$.
To establish \eqref{eq: central limit Poisson}, it is sufficient to check the Lindeberg's condition, that is, for any constant $\bar{\epsilon}>0$,
\begin{equation}
\lim_{n\to\infty}\sum_{i=1}^{n}\E\left(W_i^2 \mathbf{1}\left\{\left|W_i\right|\geq \bar{\epsilon}\right\}\right)=0.
\label{eq: lin condition Poisson}
\end{equation}
Now
\begin{eqnarray*}
    \sum_{i=1}^{n}\E\left(W_i^2\mathbf{1}\left\{|W_i|\geq \bar{\epsilon} \right\}\mid X\right) & \leq & \frac{1}{n^2}\sum_{i=1}^{n}\frac{\left(\widehat{u}^{\intercal}X_{i\cdot}\right)^2e^{-X_{i\cdot}^{\intercal}\beta}}{\rm V_{x_*}}\E\left(\frac{\epsilon_i^2}{e^{X_{i\cdot}^{\intercal}\beta}}\mathbf{1}\left\{|\epsilon_i|\geq \frac{\bar{\epsilon}n\sqrt{\rm V_{x_*}}e^{X_{i\cdot}\beta}}{\eta_n\|x_*\|_2}\right\}\right) \\
    & \leq & \max_{1\leq i \leq n}\E\left(\frac{\epsilon_i^2}{e^{X_{i\cdot}^{\intercal}\beta}}\mathbf{1}\left\{|\epsilon_i|\geq \frac{\bar{\epsilon}n\sqrt{\rm V_{x_*}}e^{X_{i\cdot}\beta}}{\eta_n\|x_*\|_2}\right\}\right) \\
    & \lesssim & \left(\frac{\bar{\epsilon}n\sqrt{\rm V_{x_*}}}{\|x_*\|_2\eta_n}\right)^{-\nu} \label{eq: verification of lin condition Poisson}
\end{eqnarray*}
The first inequality follows from the constraint $\|X\widehat{u}\|_{\infty} \lesssim \eta_n\|x_*\|_2$ in \eqref{eq: constraint 3 Poisson} while the second inequality follows from the definition of ${\rm V}_{x_*}$ in \eqref{eq: asymptotic variance}. The last inequality holds due to the assumption $\E\left[|\epsilon_i|^{2+\nu}\mid X\right] \leq \widetilde{C}$ and $\max_{1\leq i \leq n}\left|X_{i\cdot}^{\intercal}\beta\right| \leq C$ in {\rm (C1)}.

\subsection{Proof of Proposition \ref{prop: testing}}
Define $M_{\xnew} := \widehat{u}^{\intercal}\frac{1}{n}\sum_{i=1}^{n}e^{-X_{i\cdot}^{\intercal}\beta}X_{i\cdot}\epsilon_i$ and
$$
B_{\xnew} := \widehat{u}^{\intercal}\frac{1}{n}\sum_{i=1}^{n}\left(e^{-X_{i\cdot}^{\intercal}\widehat{\beta}} - e^{-X_{i\cdot}^{\intercal}\beta}\right)X_{i\cdot}\epsilon_i + \left(\xnew - \widehat{\Sigma}\widehat{u}\right)^{\intercal}(\widehat{\beta}-\beta)  + \widehat{u}^{\intercal}\frac{1}{n}\sum_{i=1}^{n}e^{-X_{i\cdot}^{\intercal}\widehat{\beta}}X_{i\cdot}\Delta_i
$$
so that $\widehat{\xnew^{\intercal}\beta} = \xnew^{\intercal}\beta + M_{\xnew} + B_{\xnew}$. When $k\log p << \sqrt{n}$ we have by Lemma \ref{lem: variance lower bound}, Propositions \ref{prop: decomposition Poisson}, \ref{prop: bias control} and \ref{prop: dominating variance}
\begin{equation}
    \frac{\widehat{\mathrm{V}}_{\xnew}}{\mathrm{V}_{\xnew}} - 1 = o_{P}(1), \; \frac{B_{x_*}}{\sqrt{\mathrm{V}_{\xnew}}} = o_P(1), \;  \frac{M_{\xnew}}{\sqrt{\mathrm{V}_{\xnew}}} \longrightarrow N(0,1)  
    \label{eq: implications}
\end{equation}
Note that
\begin{eqnarray*}
    \PP_{\theta}\left[\phi_{\xnew}(\alpha) = 1\right] & = & \PP_{\theta}\left[\left|\widehat{x_*^{\intercal}\beta}\right| \geq z_{\alpha/2}\sqrt{\widehat{\mathrm{V}}_{\xnew}}\right] \\
    & = & \PP_{\theta}\left[\widehat{x_*^{\intercal}\beta} \geq z_{\alpha/2}\sqrt{\widehat{V}_{x_*}}\right] + \PP_{\theta}\left[\widehat{x_*^{\intercal}\beta} \leq -z_{\alpha/2}\sqrt{\widehat{V}_{x_*}}\right] \\
    & = & \PP_{\theta}\left[\frac{M_{\xnew}}{\sqrt{\mathrm{V}_{\xnew}}} \geq z_{\alpha/2}\sqrt{\frac{\widehat{\mathrm{V}_{\xnew}}}{\mathrm{V}_{\xnew}}} - \frac{B_{\xnew}}{\sqrt{\mathrm{V}_{\xnew}}} - \frac{\xnew^{\intercal}\beta}{\sqrt{\mathrm{V}_{\xnew}}}\right] \\
    & + & \PP_{\theta}\left[\frac{M_{\xnew}}{\sqrt{\mathrm{V}_{\xnew}}} \leq -z_{\alpha/2}\sqrt{\frac{\widehat{\mathrm{V}_{\xnew}}}{\mathrm{V}_{\xnew}}} - \frac{B_{\xnew}}{\sqrt{\mathrm{V}_{\xnew}}} - \frac{\xnew^{\intercal}\beta}{\sqrt{\mathrm{V}_{\xnew}}}\right]
\end{eqnarray*}
Combining with \eqref{eq: implications} we can control type I error and establish the asymptotic power in \eqref{eq: power} by the definition $\xnew^{\intercal}\beta = \frac{t\|x_*\|_2}{\sqrt{n}}$.

\subsection{Proof of Theorem \ref{thm: limiting distribution QF}}
By Taylor expansion, the error $\widehat{Q}_A - Q_A$ is decomposed as $\widehat{Q}_A - Q_A = M_A + B_A$ where $$M_A := \beta_{\G}^{\intercal}(\widehat{A}-A)\beta_\G + 2\frac{1}{n}\widehat{u}_A^{\intercal}\sum_{i=1}^{n}e^{-X_{i\cdot}^{\intercal}\beta}X_{i\cdot}\epsilon_i$$ and 
\begin{eqnarray*}
    B_A := 2\left((\widehat{\beta}_\G^{\intercal}\widehat{A}\quad \mathbf{0})^{\intercal} - \widehat{\Sigma}\widehat{u}_A\right)^{\intercal}\left(\widehat{\beta}-\beta\right) & + & \left(\widehat{\beta}_\G - \beta_\G\right)^{\intercal}\widehat{A}\left(\widehat{\beta}_\G - \beta_\G\right) + \frac{2}{n}\widehat{u}_A^{\intercal}\sum_{i=1}^{n}e^{-X_{i\cdot}^{\intercal}\widehat{\beta}}X_{i\cdot}\Delta_i \\
    & + & 2\frac{1}{n}\widehat{u}_A^{\intercal}\sum_{i=1}^{n}\left(e^{-X_{i\cdot}^{\intercal}\beta}-e^{-X_{i\cdot}^{\intercal}\beta}\right)X_{i\cdot}\epsilon_i
\end{eqnarray*}

\begin{proposition}
    Suppose that Conditions \textrm{(C1)}, \eqref{eq: est property} and \textrm{(C2)} hold, then with probability larger than $1 - p^{-c} -g(n) - e^{-\sqrt{n}}$
    \begin{eqnarray}
        \left|\left((\widehat{\beta}_\G^{\intercal}\widehat{A}\quad \mathbf{0})^{\intercal} - \widehat{A}\widehat{u}_A\right)^{\intercal}\left(\widehat{\beta}-\beta\right)\right| &\lesssim &\|\widehat{A}\widehat{\beta}_\G\|_2\frac{k \log p}{n} \label{eq: bound 1 Q} \\
        \left|\left(\widehat{\beta}_\G - \beta_\G\right)^{\intercal}\widehat{A}\left(\widehat{\beta}_\G - \beta_\G\right)\right| &\lesssim& \|A\|_2\frac{k \log p}{n} \label{eq: bound 2 Q} \\
        \left|\frac{2}{n}\widehat{u}_A^{\intercal}\sum_{i=1}^{n}e^{-X_{i\cdot}^{\intercal}\widehat{\beta}}X_{i\cdot}\Delta_i\right| &\lesssim& \eta_n\|\widehat{A}\widehat{\beta}_\G\|_2\frac{k\log p}{n} 
        \label{eq: bound 3 Q} \\
        \left|\frac{1}{n}\widehat{u}_A^{\intercal}\sum_{i=1}^{n}\left(e^{-X_{i\cdot}^{\intercal}\widehat{\beta}}-e^{-X_{i\cdot}^{\intercal}\beta}\right)X_{i\cdot}\epsilon_i\right| &\lesssim& \eta_n\|\widehat{A}\widehat{\beta}_\G\|_2\frac{k \log p \log n}{n}
        \label{eq: bound 4 Q} \\
        \sqrt{\frac{1}{n}\widehat{u}_A^{\intercal}\sum_{i=1}^{n}e^{-X_{i\cdot}^{\intercal}\beta}X_{i\cdot}X_{i\cdot}^{\intercal}\widehat{u}_A} & \asymp & \|\widehat{A}\widehat{\beta}_\G\|_2 \label{eq: variance lower bound S}
    \end{eqnarray}
    \label{prop: error bound 1 Q}
\end{proposition}
 \underline{\bf Proof} : 
    The error bounds \eqref{eq: bound 1 Q} follows from Holder's inequality and \eqref{eq: projection S}. \eqref{eq: bound 3 Q}, \eqref{eq: bound 4 Q} and the lower bound \eqref{eq: variance lower bound S} follow from \eqref{eq: error bound 3 Poisson}, \eqref{eq: error bound 2 Poisson} and \eqref{eq: variance lower bound} respectively with $\xnew = \left(\widehat{\beta}_\G^{\intercal}\widehat{A} \quad \mathbf{0}\right)^{\intercal}$ and the sample splitting assumption {\rm (C2)}. The proof of \eqref{eq: bound 2 Q} follows from Lemma 10 of \cite{cai2018semi}, specifically the definition of the event $\mathcal{G}_6\left(\widehat{\beta}_\G-\beta_\G, \widehat{\beta}_\G-\beta_\G, \sqrt{n}\right)$ and hence with probability larger than $1-e^{-\sqrt{n}}$,
$$
\left|\left(\widehat{\beta}_\G-\beta_\G\right)^{\intercal} \widehat{A}\left(\widehat{\beta}_\G-\beta_\G\right)\right| \lesssim\left|\left(\widehat{\beta}_\G-\beta_\G\right)^{\intercal} A\left(\widehat{\beta}_\G-\beta_\G\right)\right| \leq\left\|A\right\|_2\left\|\widehat{\beta}_\G-\beta_\G\right\|_2^2 .
$$ 
Then use \eqref{eq: est property}.
 
Therefore we proved \eqref{eq: limiting distribution QF}. We are now proving the asymptotic normality of $M_A$.
    Note that 
    \begin{equation}
        \frac{\widehat{u}_A^{\intercal}\frac{1}{n}\sum_{i=1}^{n}e^{-X_{i\cdot}^{\intercal}\beta}X_{i\cdot}\epsilon_i}{\sqrt{\frac{1}{n^2}\widehat{u}_A^{\intercal}\sum_{i=1}^{n}e^{-X_{i\cdot}^{\intercal}\beta}X_{i\cdot}X_{i\cdot}^{\intercal}\widehat{u}_A}} \mid X,\widehat{u} \longrightarrow N(0,1)
        \label{eq: normality 1QF}
    \end{equation}
    This follows from \eqref{eq: central limit Poisson} with $\xnew := \left(\widehat{\beta}_\G^{\intercal}\widehat{A} \quad \mathbf{0}\right)^{\intercal}$ and the sample splitting assumption in {\rm (C2)}. Observe that
    \begin{equation*}
        \beta_\G^{\intercal}\left(\widehat{A}-A\right)\beta_\G =
        \begin{cases}
            0 & \text{if } A \text{ is known} \\
            \beta_\G^{\intercal}\left(\widehat{\Sigma}_{\G,\G} - \Sigma_{\G,\G}\right)\beta_\G & \text{if } A = \Sigma_{\G,\G} \text{ is unknown} 
        \end{cases}
    \end{equation*}
    and \begin{equation}
        \frac{\beta_\G^{\intercal}\left(\widehat{\Sigma}_{\G,\G} - \Sigma_{\G,\G}\right)\beta_\G}{\sqrt{\frac{1}{n} \mathbb{E}\left(\beta_\G^{\intercal} X_{i\G} X_{i\G}^{\intercal} \beta_\G-\beta_\G^{\intercal} \Sigma_{\G,\G} \beta_\G\right)^2}} \longrightarrow N(0,1)
        \label{eq: normality 2QF}
    \end{equation}
    \eqref{eq: normality 2QF} can be proved using the central limit theorem. 
    
    The proof concludes by computing the characteristic function of $M_A/\sqrt{{\rm V}_A}$. 
\noindent The proposed CI in \eqref{eq: CI QF} can be justified by the fact that $\|\widehat{A}\|_2 \frac{k \log p}{n} << \frac{\tau}{\sqrt{n}}$ and $\|\widehat{A}\widehat{\beta}_\G\|_2\frac{k \log p}{n} << \sqrt{\frac{1}{n^2}\widehat{u}_A^{\intercal}\sum_{i=1}^{n}e^{-X_{i\cdot}^{\intercal}\beta}X_{i\cdot}X_{i\cdot}^{\intercal}\widehat{u}_A}$ due to the condition $k << \sqrt{n}/\log p$.

\subsection{Proof of Proposition \ref{prop: testing QF}}

By $k \log p \ll \sqrt{n}$, we have
$$
d_n(\tau)=\frac{z_{\alpha} \sqrt{\widehat{\mathrm{V}}_{A}(\tau)}-\mathrm{B}_{A}}{z_{
\alpha} \sqrt{\mathrm{V}_{A}}}-1=o_p(1)
$$
Note that
$$
\mathbb{P}_\theta\left(\phi_A(\tau)=1\right)=\mathbb{P}_\theta\left(\widehat{Q}_A \geq z_{\alpha} \sqrt{\widehat{\mathrm{V}}_A(\tau)}\right)=\mathbb{P}_\theta\left(Q_A+M_A+B_A \geq z_{\alpha} \sqrt{\widehat{\mathrm{V}}_A(\tau)}\right) .
$$
Together with the definition of $d_n(\tau)$, we can further control the above probability by
$$
\mathbb{P}_\theta\left(M_A \geq\left(1+d_n(\tau)\right) z_{\alpha} \sqrt{\mathrm{V}_A}-Q_A\right) = \mathbb{P}_\theta\left(\frac{M_A}{\sqrt{\mathrm{V}_A}} \geq\left(1+d_n(\tau)\right) z_{\alpha}-\frac{Q_A}{\sqrt{\mathrm{V}_A}}\right) .
$$
Then we control the type I error in Proposition \ref{prop: testing QF}, following from the limiting distribution established in \eqref{eq: limiting distribution QF}. We can also establish the lower bound for the asymptotic power in Proposition \ref{prop: testing QF}. 

\subsection{Proof of Theorem \ref{thm: limiting distribution mediation}}
The proposed estimator $\widehat{\gamma^{\intercal}\beta_0}$ has the following error decomposition
\begin{eqnarray}
    \widehat{\gamma^{\intercal}\beta_0} - \gamma^{\intercal}\beta_0 = \left(\widehat{\gamma} - \gamma\right)^{\intercal}\beta_0 & + & \frac{1}{n}\widehat{u}_{\gamma}^{\intercal}\sum_{i=1}^{n}e^{-X_{i\cdot}^{\intercal}\widehat{\beta}}X_{i\cdot}\epsilon_{1 i} \\
    & + & \left((\widehat{\gamma}^{\intercal}, 0)^{\intercal} - \widehat{\Sigma}\widehat{u}_{\gamma}\right)^{\intercal}\left(\widehat{\beta} - \beta\right) \\
    & + & \frac{1}{2n}\widehat{u}_{\gamma}^{\intercal}\sum_{i=1}^{n}\frac{e^{X_{i\cdot}^{\intercal}\widetilde{\beta}}}{e^{X_{i\cdot}^{\intercal}\widehat{\beta}}}X_{i\cdot}\left(X_{i\cdot}^{\intercal}(\widehat{\beta}-\beta)\right)^{2}\\
    \label{eq: error decomposition mediation}
\end{eqnarray}
where $\widetilde{\beta}$ lies in between $\widehat{\beta}$ and $\beta$. Define
$$
M_{\gamma} := \left(\widehat{\gamma} - \gamma\right)^{\intercal}\beta_0 + \frac{1}{n}\widehat{u}_{\gamma}^{\intercal}\sum_{i=1}^{n}e^{-X_{i\cdot}^{\intercal}\beta}X_{i\cdot}\epsilon_{1 i}
$$
and 
\begin{eqnarray*}
    B_{\gamma} :=  \left((\widehat{\gamma}^{\intercal}, 0)^{\intercal} - \widehat{\Sigma}\widehat{u}_{\gamma}\right)^{\intercal}\left(\widehat{\beta} - \beta\right) & + & \frac{1}{2n}\widehat{u}_{\gamma}^{\intercal}\sum_{i=1}^{n}\frac{e^{X_{i\cdot}^{\intercal}\widetilde{\beta}}}{e^{X_{i\cdot}^{\intercal}\widehat{\beta}}}X_{i\cdot}\left(X_{i\cdot}^{\intercal}(\widehat{\beta}-\beta)\right)^{2} \\
    & + & \frac{1}{n}\widehat{u}^{\intercal}\sum_{i=1}^{n}\left(e^{-X_{i\cdot}^{\intercal}\widehat{\beta}}-e^{-X_{i\cdot}^{\intercal}\beta}\right)X_{i\cdot}\epsilon_{1 i}
\end{eqnarray*}
so that $\widehat{\gamma^{\intercal}\beta_0} = \gamma^{\intercal}\beta_0 + M_{\gamma} + B_{\gamma}$. The following Proposition bounds the bias component $B_{\gamma}$.
\begin{proposition}
    Suppose Conditions \textrm{(C1)}, \textrm{(A2)} and \eqref{eq: est property} hold. Then with probability larger than $1-p^{-c}-g(n)$ for some positive constant $c>0$,
    \begin{eqnarray}
        \left|\left((\widehat{\gamma}^{\intercal}, 0)^{\intercal} - \widehat{\Sigma}\widehat{u}_{\gamma}\right)^{\intercal}\left(\widehat{\beta} - \beta\right)\right| & \lesssim & \|\widehat{\gamma}\|_2\frac{k \log p}{n} \\
        \left|\frac{1}{n}\widehat{u}_{\gamma}^{\intercal}\sum_{i=1}^{n}e^{X_{i\cdot}^{\intercal}(\widetilde{\beta}-\widehat{\beta})}X_{i\cdot}\left(X_{i\cdot}^{\intercal}(\widehat{\beta}-\beta)\right)^{2}\right| & \lesssim & \eta_n\|\widehat{\gamma}\|_2\frac{k\log p}{n} \\
        \left|\frac{1}{n}\widehat{u}^{\intercal}\sum_{i=1}^{n}\left(e^{-X_{i\cdot}^{\intercal}\widehat{\beta}}-e^{-X_{i\cdot}^{\intercal}\beta}\right)X_{i\cdot}\epsilon_{1 i}\right| & \lesssim & \eta_n\|\widehat{\gamma}\|_2\frac{k\log p \log n}{n} \\
        \sqrt{\frac{1}{n}\widehat{u}_{\gamma}^{\intercal}\sum_{i=1}^{n}e^{-X_{i\cdot}^{\intercal}\beta}X_{i\cdot}X_{i\cdot}^{\intercal}\widehat{u}_{\gamma}} & \asymp & \|\widehat{\gamma}\|_2
    \end{eqnarray}
    \label{prop: error bound mediation}
\end{proposition}
The proof follows from Propositions \ref{prop: decomposition Poisson}, \ref{prop: bias control} and \eqref{eq: variance lower bound} using $\xnew = \left(\gamma^{\intercal}\quad \mathbf{0}\right)^{\intercal}$.
Combined with \eqref{eq: error decomposition mediation} we need to establish the asymptotic normality of $M_{\gamma}$ to prove \eqref{eq: limiting distribution of indirect effect}.
\begin{proposition}
    Suppose Conditions \textrm{(C1)} and \eqref{eq: est property} hold. Then
    \begin{equation}
        \frac{M_{\gamma}}{\sqrt{\mathrm{V}_{\gamma}^0}} \longrightarrow N(0,1)
        \label{eq: variance comp med}
    \end{equation}
    with $\mathrm{V}_{\gamma} = \frac{1}{n^2}\widehat{u}_{\gamma}^{\intercal}\sum_{i=1}^{n}e^{-X_{i\cdot}^{\intercal}\beta}X_{i\cdot}X_{i\cdot}^{\intercal}\widehat{u}_{\gamma} + \sigma_2^{2}\mathbb{E}\left(\frac{1}{\sum_{i=1}^{n}T_i^{2}}\right)$
    \label{prop: normality med}
\end{proposition}
 \underline{\bf Proof} : 
    Note that $M_{\gamma} := \left(\widehat{\gamma} - \gamma\right)^{\intercal}\beta_0 + \frac{1}{n}\widehat{u}_{\gamma}^{\intercal}\sum_{i=1}^{n}e^{-X_{i\cdot}^{\intercal}\beta}X_{i\cdot}\epsilon_{1 i}$. Again,
    \begin{eqnarray}
        \frac{\frac{1}{n}\widehat{u}_{\gamma}^{\intercal}\sum_{i=1}^{n}e^{-X_{i\cdot}^{\intercal}\beta}X_{i\cdot}\epsilon_{1 i}}{\sqrt{\frac{1}{n^2}\widehat{u}_{\gamma}^{\intercal}\sum_{i=1}^{n}e^{-X_{i\cdot}^{\intercal}\beta}X_{i\cdot}X_{i\cdot}^{\intercal}\widehat{u}_{\gamma}}} \mid X &\longrightarrow& N(0,1)
        \label{eq: normality 1 med} \\
        \frac{\left(\widehat{\gamma} - \gamma\right)^{\intercal}\beta_0}{\sqrt{\sigma_2^2\mathbb{E}\left(\frac{1}{\sum_{i=1}^{n}T_i^2}\right)}} &\longrightarrow& N(0,1)
        \label{eq: normality 2 med} \\
    \end{eqnarray}
    \eqref{eq: normality 2 med} can be proved using the central limit theorem while the limiting distribution in \eqref{eq: normality 1 med} follows from \eqref{eq: central limit Poisson} with $\xnew := \left(\widehat{\gamma}^{\intercal} \quad \mathbf{0}\right)^{\intercal}$. Then \eqref{eq: variance comp med} follows by computing the characteristic function of $M_{\gamma}/\sqrt{{\rm V}_{\gamma}}$.
 
\noindent Then \eqref{eq: limiting distribution of indirect effect} follows from Propositions \ref{prop: error bound mediation} and \ref{prop: normality med} along with the condition $k << \sqrt{n}/\log p$.

\end{document}